%% file: main.tex
\documentclass{article}
\usepackage{iclr2026_conference,times}

\usepackage{graphicx}
\usepackage{subcaption}
\usepackage{svg}
\usepackage[table]{xcolor}
\usepackage{tikz}
\usetikzlibrary{shapes.geometric, arrows.meta, arrows, positioning, fit, backgrounds, shadows, shadows.blur, matrix, calc}
\usepackage{colortbl}
\usepackage{tabularx}
\usepackage{booktabs}
\usepackage{placeins}
\usepackage{float}
\usepackage{hyperref}
\usepackage{url}
\input{math_commands.tex}

\iclrfinalcopy 

\begin{document}

\title{Accelerating Materials Discovery: \\ 
Learning a Universal Representation of \\ 
Chemical Processes for Cross-Domain \\ 
Property Prediction} 

\author{Mikhail Tsitsvero\thanks{Correspondence to: \texttt{m.tsitsvero@crowdchem.net}, \texttt{a.nakao@crowdchem.net}.} \\
CrowdChem, Inc., Tokyo, Japan
\And
Atsuyuki Nakao\footnotemark[1] \\
CrowdChem, Inc., Tokyo, Japan 
\And
Hisaki Ikebata \\
CrowdChem, Inc., Tokyo, Japan
}

\maketitle
\lhead{} 

\begin{abstract} 
Experimental validation of chemical processes is slow and costly, limiting exploration in materials discovery. Machine learning can prioritize promising candidates, but existing data in patents and literature is heterogeneous and difficult to use. We introduce a universal directed-tree process-graph representation that unifies unstructured text, molecular structures, and numeric measurements into a single machine-readable format. To learn from this structured data, we developed a multi-modal graph neural network with a property-conditioned attention mechanism. Trained on approximately 700,000 process graphs from nearly 9,000 diverse documents, our model learns semantically rich embeddings that generalize across domains. When fine-tuned on compact, domain-specific datasets, the pretrained model achieves strong performance, demonstrating that universal process representations learned at scale transfer effectively to specialized prediction tasks with minimal additional data.
\end{abstract}

\section{Introduction}
Experimental validation of chemical processes is slow and costly, which limits throughput and constrains exploration of the vast design space in materials discovery. Data-driven methods can mitigate this bottleneck by learning from prior experiments to prioritize promising materials and conditions, thereby focusing scarce laboratory effort on fewer, more informative trials. Realizing this potential requires machine-readable process representations that unify heterogeneous data modalities—unstructured text, tabular measurements, and molecular structures—into a single representation. Efficiently representing chemical processes is a long-standing challenge. Traditional methods often struggle to capture the complex interplay of materials, conditions, and steps in a unified format. Furthermore, because processes vary widely in complexity—from simple mixtures to multi-stage syntheses—they produce data of variable size that is difficult to encode into fixed-length vectors without semantic loss. This lack of a universal representation limits the application of machine learning. Without a shared vocabulary of substructures reused across examples, models cannot easily learn features that transfer across domains. 

Recent advancements in multi-modal AI have begun to address these challenges by integrating diverse data types such as molecular structures, spectra, and images, enabling more comprehensive analyses and predictions in chemistry and materials science. For instance, recent models demonstrate progress towards large multi-modal frameworks capable of handling multiple chemical modalities~\cite{zhao2024chemdfm,HandokoMade2025AIGenerativeMaterialsReview}. Moreover, the emergence of large language models is poised to revolutionize various aspects of materials science research~\cite{lei2024materials}. Understanding which elements of a complex chemical process are most influential for the final properties is crucial for process optimization.   
In parallel, self-driving laboratory platforms that couple AI-driven experiment planning with automated execution are enabling closed-loop discovery and more sample-efficient exploration of processing and formulation spaces~\cite{tom2024selfdriving,Tobias2025,Epps2021Universal}.

In this work, we propose a graph-based representation of chemical processes and a multi-modal GNN-based model with an attention mechanism that learns from it. Our contributions are: (i) a universal directed-tree process representation that jointly encodes experimental conditions, molecular structures, and stepwise operations; (ii) a multi-modal, multi-task GNN that learns a shared process embedding and predicts properties via task-specific output heads; and (iii) a flexible fine-tuning framework that effectively adapts the pretrained backbone to specific application domains. Large-scale pretraining over tens of thousands of (document+property) tasks is used to learn the diversity of chemical processes and to produce a transferable process embedding. We demonstrate that this shared representation allows for data-efficient adaptation to new domains using strategies ranging from lightweight head calibration and residual adaptors to full-model fine-tuning, and enabling accurate property prediction even on smaller, specialized datasets. Here, each task corresponds to a (document+property) pair; the latent representation is shared across tasks, while the output heads are trained per task. This design enables strong performance after fine-tuning and can be integrated into active learning loops for autonomous experimentation. The remainder of this paper is organized as follows. Section~\ref{sec:representation} formalizes the directed-tree representation for chemical processes and the associated node and edge types. Section~\ref{sec:model} presents the multi-modal graph neural network that operates on this representation and outlines the training and fine-tuning setup. Section~\ref{sec:finetuning} details the application-focused fine-tuning methodology and an illustrative case study. Section~\ref{sec:future_directions} discusses future directions and multi-modal extensions. Section~\ref{sec:conclusion} concludes with limitations and a summary.

\section{The Directed-Tree Representation for Chemical Processes}
\label{sec:representation}

\begin{figure}[H]
  \centering
  \includegraphics[width=0.95\linewidth]{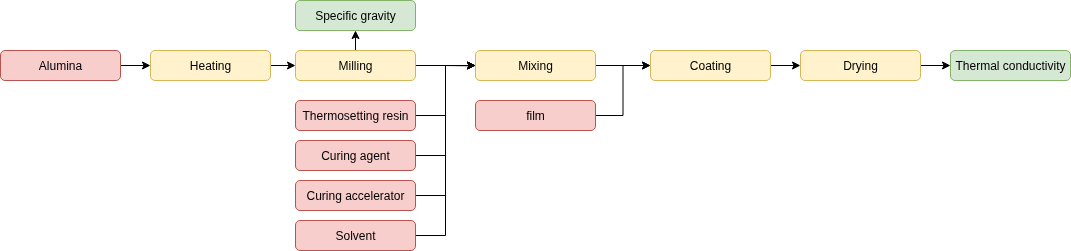}
  \caption{Experimental process for a resin composite containing alumina filler. The alumina is pretreated through heat treatment and milling. After that, the specific gravity of the alumina is measured. Subsequently, a slurry mixture of resin, curing agent, curing accelerator, solvent, and filler is coated onto a film, dried, and the thermal conductivity of the cured resin composite is measured.}
  \label{fig:process_diagram}
\end{figure}

\begin{figure}[htbp]  
  \centering
  \includegraphics[width=0.9\linewidth]{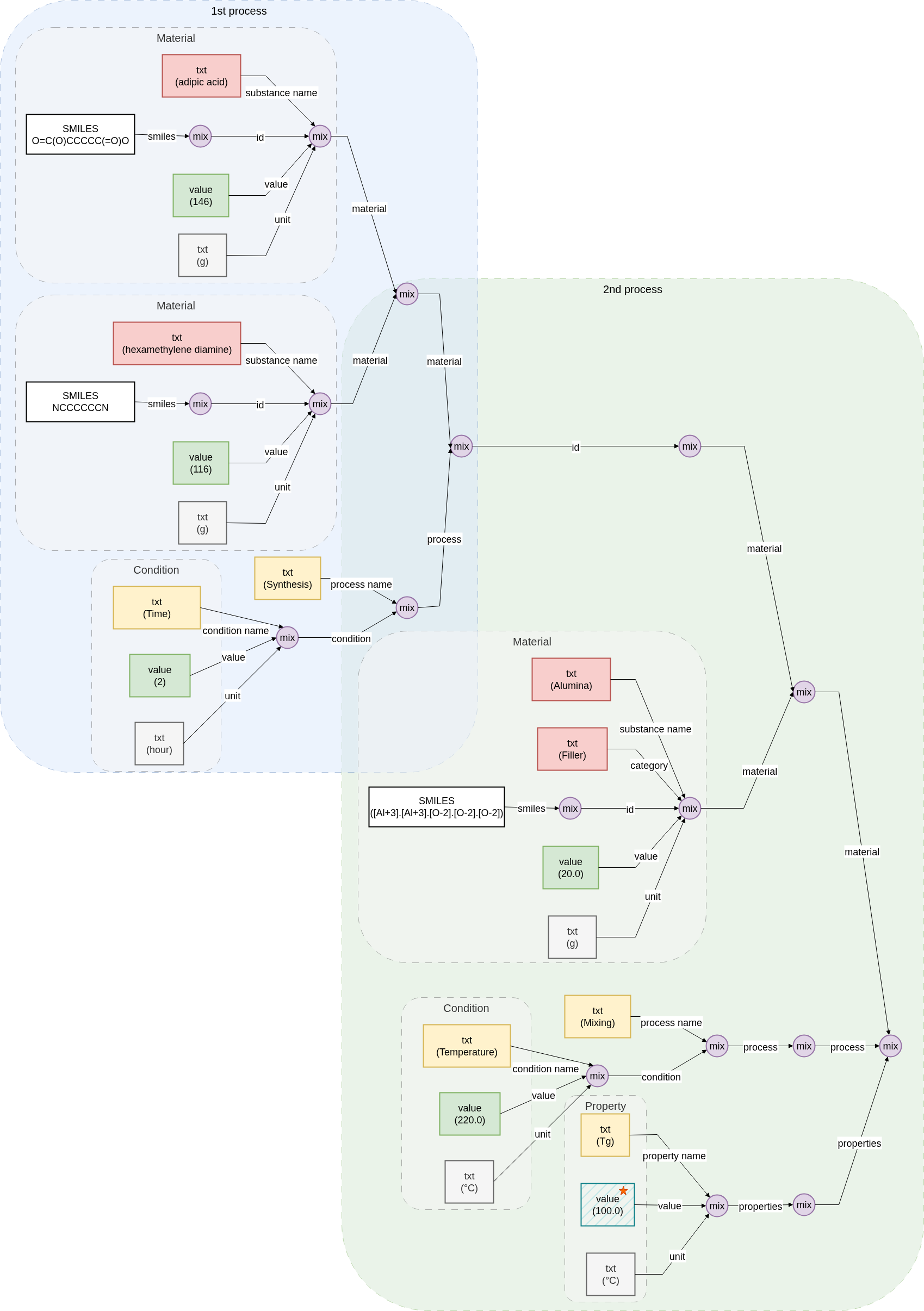}
  \caption{Two-step chemical process encoded as our directed-tree input. \emph{In the \textbf{1st process} (blue region), we \textbf{synthesize} a polyamide by combining \textbf{adipic acid} (SMILES \texttt{O=C(O)CCCCCC(=O)O}, \emph{146 g}) with \textbf{hexamethylene diamine} (SMILES \texttt{NCCCCCCN}, \emph{116 g}) in a \textbf{synthesis} step lasting \emph{2 h}. In the \textbf{2nd process} (green region), the resulting product is \textbf{mixed} with \textbf{Alumina} (\textbf{Filler}, \emph{20.0 g}) in a \textbf{mixing} step at \emph{220.0\,$^\circ$C}, and the target property, glass-transition temperature \textbf{Tg}, is recorded as \emph{100.0\,$^\circ$C}.} Node types (\texttt{mix} for structural grouping, \texttt{txt}, \texttt{value}, \texttt{SMILES}) and labeled edges (\texttt{material}, \texttt{process}, \texttt{condition}, \texttt{properties}, \texttt{id}) define the directed tree used by our model; the output product subtree of the first process is linked into the second process via an \texttt{id} edge from the downstream material node.}
  \label{fig:process_representation} 
\end{figure}

Before turning to our graph-based representation, it is helpful to visualize a typical experimental workflow in the more familiar flowchart style. As an example, Figure~\ref{fig:process_diagram} illustrates the experimental process for a resin composite containing alumina filler. The alumina is pretreated through heat treatment and milling. After that, the specific gravity of the alumina is measured. Subsequently, a slurry mixture of resin, curing agent, curing accelerator, solvent, and filler is coated onto a film, dried, and the thermal conductivity of the cured resin composite is measured. This conventional diagram highlights temporal ordering but treats each box in a generic way. Figure~\ref{fig:process_representation} provides a detailed view of our data representation using an example of a multi-step process that links polyamide synthesis with a subsequent mixing step. This directed-tree schema organizes the data hierarchically and makes the underlying materials, operations, conditions, and target properties explicit as nodes and labeled edges.

Conceptually, our format represents experiments as graphs based on three simple principles:
\begin{enumerate}
    \item Experiments can be viewed as flowcharts---graphs of interconnected process steps---where edges describe how the output of one step becomes the input of another.
    \item Each process contains detailed process conditions, the materials used in that step, and physical property measurements of the resulting product.
    \item Material information itself is structured and may include substance names, product names, physical properties, and chemical structures.
\end{enumerate}

Now, we propose a directed tree-graph structure to serve as a universal representation for such chemical processes. We consider this representation universal in its capacity to encode any chemical process that can be described as a sequence of operations on materials under specific conditions, a paradigm that covers the vast majority of experimental procedures found in scientific literature and patents. In such documents, going beyond simple table data is essential because key scientific information is distributed across figures, images, chemical structures, reaction schemes, and plots. To ingest this heterogeneous, multi-modal evidence, our representation is designed to be highly flexible. Each directed tree has a single root node representing the (document, property) example, from which all material, process, condition, and property subtrees descend. Each node in the graph represents a specific component of the process, such as a material, a piece of equipment, a processing step, or a measured property. The directed edges define the relationships between these components, creating a hierarchical structure that mirrors the flow of a real-world chemical experiment.

The power of this representation also lies in its ability to integrate other data types. In this representation, chemical structures are encoded by SMILES~\citep{Weininger1988} strings, which are converted to molecular graphs and processed by graph neural networks (GNNs)~\citep{pmlr-v70-gilmer17a} to produce embedding vectors attached to the corresponding process nodes. Textual information, such as procedural descriptions or material names, is converted into dense vector embeddings. Other numerical data, including physical quantities, are also transformed into a vector format. This multi-modal approach allows the model to build a comprehensive understanding of the process by combining information from all available sources into a single, unified graph structure.

\paragraph{Node and edge types used in the directed-tree input}
Our input graphs use a small, fixed set of node and edge categories (summarized in Table~\ref{tab:node_types} and Table~\ref{tab:edge_types}) that are sufficient to encode heterogeneous experimental data. Nodes fall into four primary types: \texttt{mix} (structural containers), \texttt{value} (numeric leaves), \texttt{txt} (textual leaves), and \texttt{SMILES} (molecular structure references). Two special markers, \texttt{predvalue} and \texttt{pred}, reuse the \texttt{value} and \texttt{txt} types to denote the target property value and its name, respectively. Edge types are derived directly from the curated document data and are mapped to 17 categorical edge labels. Crucially, material, condition, process, and property subtrees are expressed using the same \texttt{mix}-node patterns across all graphs, so the model sees a consistent vocabulary of features and can learn universal representations of these concepts that transfer across documents and domains. 

At the level of subtrees, each process node connects three main kinds of information:
\begin{itemize}
    \item \textbf{Material usage subtree.} A material-usage node (a \texttt{mix} container) connects to (i) a \texttt{category} node that encodes how the material is used within the process (for example, ink vs.\ film, electrode vs.\ electrolyte), (ii) an \texttt{id} edge that points to a subtree aggregating detailed material information, and (iii) \texttt{value} and \texttt{unit} nodes that record the usage quantity and its unit.
    \item \textbf{Condition subtree.} A condition subtree follows a simple pattern consisting of a \texttt{condition name} node and its associated \texttt{value} and \texttt{unit} nodes (for example, Time = 2~h, Temperature = 220~$^\circ$C).
    \item \textbf{Property subtree.} A property node aggregates measurement results. It consists of \texttt{property name}, \texttt{value}, and \texttt{unit} nodes, along with measurement conditions (represented by the conditions subtree).
    \item \textbf{Process information subtree.} A process node connects via a \texttt{process name} edge to a concise label such as ``Heating'', ``Milling'', or ``Mixing'', and via \texttt{condition} edges to condition subtrees. This subtree describes the operational parameters (name and conditions) of the step.
\end{itemize}
 
Tables~\ref{tab:node_types} and~\ref{tab:edge_types} provide further details, listing specific node types and edge labels with brief explanations and representative examples of child-node attribute values. 

Detailed material information is linked to the process subtree through the \texttt{id} edge. If the material is an intermediate produced by another process in the same experiment, the process subtree that generates that material itself serves as the detailed information. Otherwise, the material-information subtree contains edges such as \texttt{substance name} (specific material name, ranging from concrete compound names to abstract categories like silane coupling agent or film), \texttt{product name} (product model number or commercial identifier), \texttt{property} (material properties encoded in the same format as process-level properties), and \texttt{SMILES} for chemical structure information. For polymers, inorganic compounds, and mixtures, the SMILES information may consist of one or more subtrees: low-molecular-weight compounds and inorganic compositions typically have a single SMILES entry (for example, \texttt{[OH-].[Na+]}), whereas polymers can have multiple repeating-unit subtrees whose SMILES strings use \texttt{[*]} at the bonding positions and whose relative fractions are recorded in ratio fields. Terminal-group information for polymers is encoded as a separate SMILES subtree linked via a dedicated terminal-SMILES edge. A free-text \texttt{memo} node can also be attached to record information that does not fit elsewhere (such as polymer bonding patterns).

To illustrate the directed-tree representation, Figure~\ref{fig:process_representation} presents a concrete example with two linked processes: a polyamide synthesis followed by a mixing step. The diagram explicitly maps the experimental flow to our graph structure. Material and condition subtrees are grouped by \texttt{mix} nodes and connected via \emph{material} and \emph{condition} edges, while the resulting mixture from the first process becomes an input to the second. A properties subtree attaches the measured target (e.g., glass-transition temperature) as a \texttt{predvalue} node. When the curated data contains lists of items, our compiler inserts additional \texttt{mix} nodes that serve as uniform list containers, allowing singletons and multi-entry lists to share the same schema and simplifying downstream message passing. Together, these linked processes demonstrate how heterogeneous experimental information is normalized into a single directed-tree graph. 

\begin{table}[t] 
  \centering
  \caption{Node types used in the process graphs.}  
  \label{tab:node_types}
  {\small\setlength{\arrayrulewidth}{0.2pt}\arrayrulecolor{gray!30}%
  \begin{tabularx}{\linewidth}{l >{\raggedright\arraybackslash}X >{\raggedright\arraybackslash}X >{\raggedright\arraybackslash}X}
  \toprule 
  \textbf{Name} & \textbf{Internal Attributes} & \textbf{Explanation} & \textbf{Example attribute value} \\ 
  \midrule
  \texttt{mix} & None & Structural container node (grouping) used to group children; created for dictionaries/lists and does not store a numeric, text, or SMILES attribute itself. & --- (no attribute value; acts only as a container for child nodes). \\
  \hline
  \texttt{value} & \texttt{value} & Numeric leaf node that stores a scalar quantity; ordinal scales are translated when needed. & 20.0 (g of filler), 220.0 (temperature), 2.0 (time in hours). \\
  \hline
  \texttt{predvalue} & \texttt{value} (marked as target) & Numeric value node marked as the target property for supervised learning. & 100.0 (glass-transition temperature Tg used as the prediction target). \\
  \hline
  \texttt{txt} & text string & Text leaf node; the raw string is embedded into a 512-D text embedding used by the model. Note: \texttt{property name} here refers to properties recorded as input or context. & \emph{Alumina}, \emph{Tg}, \emph{thermal conductivity}, \emph{g}, \emph{Temperature}. \\
  \hline
  \texttt{pred} & text string (marked as target) & Text node containing the target property name (query) used to condition attention and readout. & \emph{Tg}, \emph{specific gravity}, \emph{thermal conductivity}. \\
  \hline
  \texttt{SMILES} & SMILES string & Molecular-structure node; the SMILES string is converted to an atom graph and encoded by a molecular GNN, and the resulting embedding vector is stored here. & \texttt{O=C(O)CCCCCC(=O)O}, \texttt{[OH-].[Na+]}, or polymer repeating-unit SMILES using \texttt{[*]} at bonding positions. \\
  \bottomrule
  \end{tabularx}%
  }\arrayrulecolor{black} 
  \end{table} 
  
  \begin{table}[htbp]
  \centering
  \caption{Edge types used in the process graphs. Each row lists an edge name in the directed-tree schema, a brief explanation of its role, and representative examples of node attribute values that typically appear at the corresponding child nodes.} 
  \label{tab:edge_types}
  {\small\setlength{\arrayrulewidth}{0.2pt}\arrayrulecolor{gray!30}\setlength{\tabcolsep}{3pt}%
  \begin{tabularx}{\linewidth}{l >{\raggedright\arraybackslash}X >{\raggedright\arraybackslash}X}
  \toprule
  \textbf{Name} & \textbf{Explanation} & \textbf{Example of node attribute value} \\
  \midrule
  \texttt{material} & Connects material-usage subtrees to the process node. & Edge from a process node to a material container that lists inputs for the step. \\
  \hline
  \texttt{category} & Role of the material within the step. & \emph{Filler}, \emph{Binder}, \emph{ink}, \emph{film}, \emph{electrode}, \emph{electrolyte}. \\
  \hline
  \texttt{id} & Pointer from a material-usage node to its detailed material-information subtree (which may itself be a process subtree for intermediates). & Edge from ``polyamide mixture'' material usage to the subtree describing its composition or upstream synthesis process. \\
  \hline
  \texttt{substance name} & Specific material name (compound name, chemical family, or abstract part such as film or printed board). & \emph{Alumina}, \emph{adipic acid}, \emph{hexamethylene diamine}, \emph{film}. \\
  \hline
  \texttt{value} & Connects to numeric value nodes (amounts, conditions, measured properties, repeating-unit ratios, etc.). & 20.0 (g of filler), 220.0 (temperature), 100.0 (Tg), 0.4 (unit ratio). \\
  \hline
  \texttt{unit} & Unit associated with a numeric value. & \emph{g}, \emph{h}, \emph{$^\circ$C}, \emph{wt\%}. \\
  \hline
  \texttt{product name} & Commercial product or model name used to distinguish materials when detailed composition is unknown. & Trade name or grade code of a purchased resin or additive. \\  
  \hline
  \texttt{SMILES} / \texttt{main SMILES} & Edge to SMILES information for the main structural description (single compound, composition formula, or collection of polymer repeating units with ratios). & \texttt{O=C(O)CCCCCC(=O)O}, \texttt{[OH-].[Na+]}, or repeating-unit SMILES with associated ratios. \\
  \hline
  \texttt{properties} & Connects to a subtree aggregating physical-property measurements for the process or material. & Subtree containing nodes for \emph{Tg}, \emph{thermal conductivity}, or \emph{specific gravity}. \\
  \hline
  \texttt{internal SMILES} & Internal entries within a SMILES container, for example individual polymer repeating units with associated ratio information. & Repeating-unit SMILES and ratio pairs in a copolymer description. \\
  \hline
  \texttt{terminal SMILES} & Edge to terminal-group SMILES for polymers (tail SMILES). & SMILES for a polymer end group such as a capping or initiator fragment. \\
  \hline
  \texttt{property name} & Name of a measured or material property. & \emph{Tg}, \emph{thermal conductivity}, \emph{specific gravity}. \\ 
  \hline
  \texttt{condition name} & Name of a process or measurement condition. & \emph{Time}, \emph{Temperature}, \emph{Mixing speed}. \\
  \hline
  \texttt{condition} & Connects to a subtree aggregating (\texttt{condition name}, \texttt{value}, \texttt{unit}) triplets describing process or measurement conditions. & Subtree representing Time~=~2~h or Temperature~=~220.0~$^\circ$C. \\ 
  \hline
  \texttt{process name} & Concise name of a process step. & \emph{Heating}, \emph{Milling}, \emph{Mixing}, \emph{Coating}, \emph{Drying}. \\
  \hline
  \texttt{process} & Connects to process subtrees that gather materials, conditions, and properties for a given step. & Edge from a higher-level ``process list'' node to an individual \emph{Mixing} or \emph{Coating} step subtree. \\
  \hline
  \texttt{memo} & Free-text memo for information not captured elsewhere (for example, polymer bonding patterns or special notes). & Note describing special polymer bonding patterns or experimental remarks that do not fit other fields. \\
  \bottomrule
  \end{tabularx}%
  }\arrayrulecolor{black}  
  \end{table} 

\paragraph{Multi-step experiments and linked processes.}
The directed-tree representation also captures multi-step workflows in a way that preserves the experimental topology. Conceptually, each process step corresponds to a distinct subtree containing input materials, an operation name, conditions, and products. When an experiment involves several chained steps, references between steps are resolved into nested structures: the product subtree of step \(k\) is injected as a material subtree feeding step \(k{+}1\). Concretely, the product of a process is represented as a subtree that aggregates all information produced by that step; the edge labeled \texttt{id} from a downstream material node then points to the root of this product subtree, so that repeating this pattern in a nested manner represents the overall experimental flow. This design lets the representation scale from one-step measurements to long synthetic routes while keeping a strictly tree-structured, schema-consistent graph for learning.

\paragraph{Data conversion workflow.}
Our directed-tree graphs are the endpoint of a staged, largely lossless conversion pipeline rather than a direct, opaque export from raw documents. We first curate heterogeneous sources (patents, reports, and lab notebooks) directly into schema-validated JSON records that explicitly label text, numeric quantities, and molecular structures (SMILES), and retain both original values and their normalized counterparts (for example, after unit translation). Finally, the JSON records are deterministically mapped into directed-tree graphs by wrapping nested dictionaries and lists into structural container nodes (referred to as mix nodes) and attaching modality-specific attributes—molecular graphs, raw numeric values, and text embeddings—at the leaves. In this framework, the molecular and numeric embeddings are learned during training, whereas text embeddings are pre-computed. This workflow separates human curation from programmatic conversion, preserves provenance at every stage, and allows processes with fundamentally different surface formats or flow diagrams to be handled collectively once expressed in the common graph representation.

In the present work, all chemical process graphs were curated by a team of chemical experts to ensure fidelity and consistency (see Appendix~\ref{app:dataset_curation}). This task was performed using a specially designed graphical user interface (GUI). At first glance, our JSON-based data processing workflow may appear complex; however, it allowed seamless integration with the data processing team as well as continuous production. Looking forward, large language models (LLMs) will be indispensable—and strictly complementary—for scaling this curation. The JSON-based intermediate representation provides a discrete, schema-validated target that is naturally compatible with the structured generation capabilities of modern LLMs, enabling them to reliably transform unstructured literature and patent text into the strict format required by our schema. This perspective aligns with recent views on LLMs in materials science~\cite{lei2024materials,foppiano2024mining,buffo2025assessment,Ansari2024Eunomia,schilling2025text} and emphasizes that LLMs augment, rather than replace, our graph-based approach.

Direct reliance on LLMs for numerical property prediction remains challenging. First, they struggle with regression tasks because numbers are processed as discrete tokens rather than continuous values, often leading to poor precision. Second, LLMs' accuracy typically declines with increasing context length, limiting the feasibility of inserting large datasets or detailed process histories directly into the context window. Finally, fine-tuning massive LLMs for specific tasks incurs high computational costs. In our workflow, therefore, LLMs are intended to assist curation by helping convert heterogeneous prose and tables into the structured fields of our directed-tree schema, whereas the core predictive modeling is performed by a multi-modal, structure-aware GNN operating on explicit process graphs. This separation leverages LLM strengths in language understanding while preserving precise, auditable graph structure and cross-modal reasoning that a pure LLM sequence approach does not natively provide.

\section{Multi-modal model for cross-domain property prediction}
\label{sec:model}
\begin{figure}[t]
  \centering
  \includegraphics[width=1.0\linewidth]{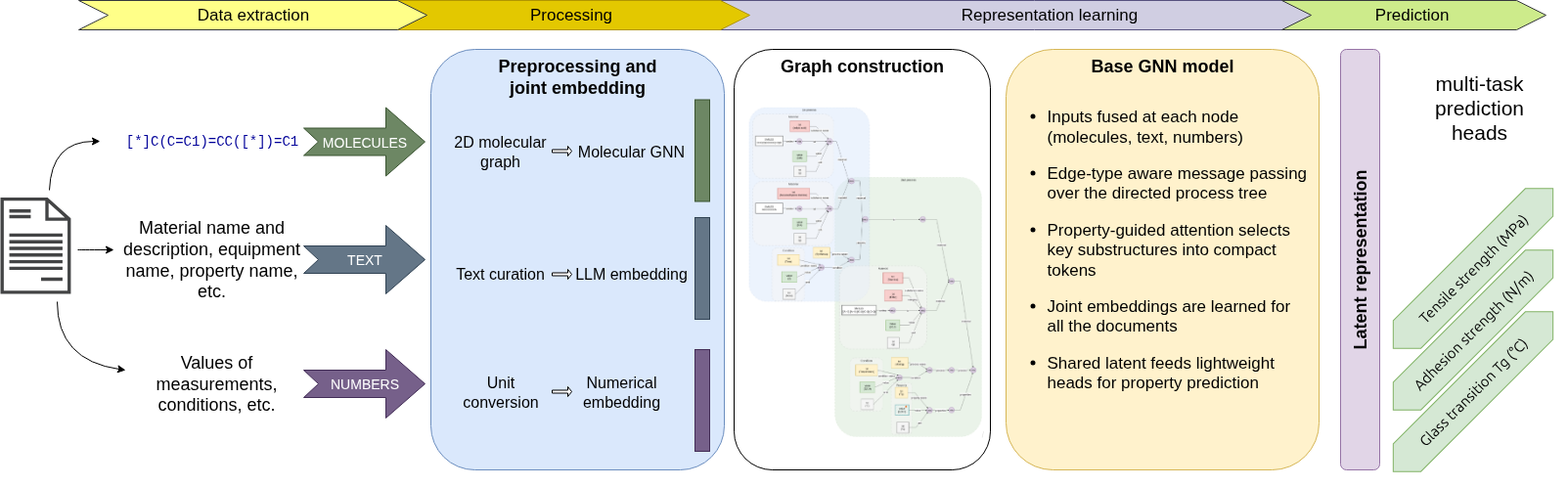} 
  \caption{The architecture of the multi-modal, multi-task graph neural network. The model takes the directed tree representation of a chemical process as input. Different node types are processed by specialized encoders: a graph neural network for molecular structures (SMILES), LLM embedding for text, and separate network for numerical values. The resulting embeddings are then passed to a main process GNN that performs message passing on the entire graph. Cross-modal attention pools information into a compact latent representation (property-conditioned tokens), which is shared across tasks. Task-specific output heads (indexed by document+property) map the shared latent to the corresponding prediction.}
  \label{fig:model_architecture}
\end{figure}

To leverage our universal representation, we have developed a graph neural network (GNN) model. The model architecture, illustrated in Figure~\ref{fig:model_architecture}, is specifically designed to handle the multi-modal and hierarchical nature of the chemical process graphs. It employs a multi-stage and directed graph approach to information processing. Initially, specialized sub-networks process different data modalities. For instance, a dedicated GNN is used to learn feature representations for chemical structures from their atomic graphs, while textual descriptions are encoded using a text embeddings by large language model.

These initial node representations are then fed into a main GNN that operates on the entire process graph. This network uses a transformer-based message passing mechanism, allowing it to effectively capture long-range dependencies and complex interactions between different parts of the chemical process. The node update rule is given by:
\begin{equation}
\mathbf{h}'_i = \mathbf{W}_1 \mathbf{h}_i + \sum_{j \in \mathcal{N}(i)} \alpha_{ij} (\mathbf{W}_2 \mathbf{h}_j + \mathbf{W}_e \mathbf{e}_{ij}),
\end{equation}
where $\mathbf{h}_i$ is the node feature, $\mathbf{e}_{ij}$ is the edge feature, $\mathbf{W}_e$ projects the edge attributes, and $\alpha_{ij}$ are attention coefficients (see Appendix~\ref{app:gnn_attention} for full details). The architecture is multi-task: a single shared backbone produces a latent representation, which is consumed by a set of lightweight task-specific output heads (one per document+property task). 

In simple terms, cross-modal attention is applied at three places, always using the target property-name embedding as the query vector $\mathbf{q}$. Crucially, the query vector $\mathbf{q}$ encodes the target property's semantic identity, while the key $\mathbf{k}$ and value $\mathbf{v}$ vectors are derived from the specific experiment's graph nodes. This allows the model to dynamically extract and aggregate information relevant to the target property from the unique material and condition configuration of each experiment. The attention is applied: (i) at the molecule level, to pool atom features into a property-conditioned molecule embedding from the SMILES graph via attention over the molecular graph; (ii) at the process level, after message passing on the entire process graph, to attend from the target property over structural (container) nodes in the directed tree and collect a small set of latent tokens that summarize the most relevant substructures; and (iii) at readout, to attend from the target property over those refined latent tokens to form a single prediction vector (see Appendix~\ref{app:prop_readout} for attention-pooling details). At stage (ii), the attention mechanism has access to enriched node representations that fuse materials (names and categories), property-conditioned molecular embeddings, measured quantities and their units, process step names, and condition names/values—all integrated through message passing. By restricting attention to structural container nodes rather than all graph nodes, the architecture leverages the hierarchical tree structure: the prior message passing has already propagated information from leaf nodes (values, text, molecular embeddings) into these containers, so attending over them provides access to the complete process context while respecting the compositional organization of the data. This hierarchical property-conditioning highlights the parts of the experiment most relevant to the requested property before producing a prediction.

\subsection{Input processing} 
Our input graphs follow the directed-tree schema from Section~\ref{sec:representation}. Each leaf node carries one of three modalities: numeric values, short text, or molecular structures (SMILES). Internal nodes serve as structural containers that preserve the hierarchy of materials, processing steps, conditions, and properties. For supervised training, the measured target value and its property name are explicitly marked, but the target value is not provided at inference time.

Before message passing, node attributes are transformed into dense representations:
- Numeric leaves are mapped to a fixed-length vector using a smooth bank of learned radial functions on a logarithmic scale, which captures sign and order-of-magnitude while remaining robust to skewed value distributions.
- Text leaves and the target property name are embedded with a pretrained language model and linearly projected to the model dimension. These embeddings provide semantic anchors for materials, units, condition names, and property descriptors.
- SMILES leaves are converted to molecular graphs and processed with a molecular graph encoder to produce atom-level features. A property-conditioned attention pooling aggregates atom features into a molecule embedding that is assigned back to the corresponding process-node location.

In addition, node types and edge labels (the categories listed in Section~\ref{sec:representation}) are encoded and supplied to the main graph network. This yields, for every node, a unified vector that fuses numeric, textual, and molecular information while preserving the original directed-tree connectivity. Concretely, the per-node numeric, text, and molecular embeddings are concatenated and then processed by a transformer-style graph convolution over the directed tree; categorical edge labels are embedded and provided as edge attributes to the convolution.

\subsection{Multi-task prediction and inference pipeline}
The model proceeds in five stages, all conditioned on the target property embedding:
1) Modality encoders transform numeric, text, and molecular inputs into node-aligned vectors. The molecular encoder runs on atom graphs and returns property-conditioned molecule embeddings, which are placed at the appropriate nodes in the process graph.
2) Graph message passing propagates information along the directed-tree using edge-aware attention and residual connections, allowing distant parts of the process to interact through their categorical relationships (material, process, condition, unit, etc.).
3) Property-conditioned pooling to latent tokens: the model computes attention from the target property onto a subset of structural (container) nodes and aggregates them into a small, fixed set of latent tokens. These tokens act as a compact "scratchpad" that is not part of the original tree but summarizes the relevant substructures for the current prediction.
4) Token refinement or compression: a short stack of property-conditioned attention layers optionally refines and/or compresses the latent tokens. In our implementation we use stacked property-conditioned selection layers; a self-attention refinement block can be inserted but is not required.
5) Property-conditioned readout and prediction: attention from the target property over the refined tokens produces a single latent vector. This shared latent is routed to a task-specific head (document+property) that maps it to a scalar prediction. 

This design cleanly separates three roles: (i) modality-specific extraction at the leaves, (ii) structure-aware propagation over the directed tree, and (iii) property-conditioned selection and aggregation into latent tokens. The latter two stages create the effect of "cross-attention to unconnected nodes" by allowing the target property to focus on any part of the process graph and to reason in a compact token space that lives alongside, but outside, the original tree.  

\paragraph{Property-conditioned readout.} Our readout is a permutation-invariant, property-conditioned attention pooling. The target property embedding acts as a query over structural nodes and produces a weighted aggregation that is mapped to a graph-level readout vector $\mathbf{z}_{\text{readout}}$. We transform this into the final latent vector used for prediction,
\[
\mathbf{z}_i \;=\; \tanh\!\big(\mathrm{LN}(\mathbf{W}_z(\mathbf{y}_{\text{prop}})\,\mathbf{z}_{\text{readout}} + \mathbf{b}_z(\mathbf{y}_{\text{prop}}))\big), 
\]
where $\mathbf{W}_z(\cdot)$ and $\mathbf{b}_z(\cdot)$ are linear projections from the target property embedding $\mathbf{y}_{\text{prop}}$, enabling the transformation to adapt to the specific property being predicted. This latent vector is then used in the final prediction layer; see \Eqref{eq:pred_modulation} (additional hyperparameters in the Supplementary Information).

\paragraph{Final prediction with task modulation.} Let $\mathbf{z}_i\in\mathbb{R}^d$ be the readout vector for example $i$ (after property-conditioned pooling), and let $\mathbf{W}_p(\mathbf{y}_{\text{prop}})\in\mathbb{R}^d$ be a base projection derived from the target property embedding via a learned linear transformation. For task $t(i)$ we learn three task parameters: an element-wise modulation vector $\mathbf{w}_{t(i)}\in\mathbb{R}^d$ and scalars $a_{t(i)}, b_{t(i)}$. The scalar prediction is
\begin{equation}
\label{eq:pred_modulation}
p_i \,=\, a_{t(i)}\, \Big(\big(\mathbf{W}_p(\mathbf{y}_{\text{prop}}) \odot (1 + \tfrac{\mathbf{w}_{t(i)}}{100})\big)^{\top} \mathbf{z}_i\Big) + b_{t(i)} ,
\end{equation}
where $\odot$ denotes element-wise multiplication. The scaling factor $1/100$ ensures the modulation starts as a small perturbation. Thus the per-task vector $\mathbf{w}_{t}$ softly rescales each latent dimension before the dot-product with $\mathbf{z}_i$, enabling lightweight task-specific adaptation while the backbone and property-conditioned projection $\mathbf{W}_p(\cdot)$ remain shared across tasks. In addition to the prediction $p_i$, the output head produces a per-task aleatoric uncertainty estimate $\sigma_{t(i)}$ that quantifies the expected noise level for task $t(i)$. Here $\mathbf{z}_i$ is obtained from the property-conditioned readout via an intermediate transformation layer $\mathbf{z}_i = \tanh(\mathrm{LN}(\mathbf{W}_z(\mathbf{y}_{\text{prop}})\mathbf{z}_{\text{readout}} + \mathbf{b}_z(\mathbf{y}_{\text{prop}})))$; see Appendix~\ref{app:adaptor_finetune_details} for complete details of the output head architecture. Training uses a heteroscedastic objective that jointly optimizes predictions and uncertainty estimates via a log-variance term and a variance-normalized squared error, plus an $\ell_2$ regularizer on the head's output-weight term (see Appendix~\ref{app:heteroscedastic_loss}).

\subsection{Model Training}   
The model was trained on a massive, curated dataset of approximately $700{,}000$ process graphs extracted from diverse sources, including patents and scientific literature (see Appendix~\ref{app:dataset_curation} for details on dataset curation and scope). Training follows a multi-task regime where each task corresponds to a unique (document, property) pair. In this framework, the backbone and latent representations are shared across all tasks to capture generalizable process structures, while the output layer parameters are task-specific. This design encourages a shared representation that captures universal chemical logic while allowing for specialization in the final prediction heads. Further details on the training procedure, computational infrastructure, and hyperparameters are provided in Appendix~\ref{app:base_training_details}. 
 


\section{Base-model fine-tuning for application-specific use cases}
\label{sec:finetuning}
The pretrained model in Section~\ref{sec:model} learns a shared representation from hundreds of thousands of process graphs and tens of thousands of (document+property) tasks. For practical deployment, organizations usually bring a smaller, domain-focused dataset (e.g., a particular product line, manufacturing route, or assay). We consider three complementary fine-tuning scenarios that adapt the pretrained backbone to such use cases. 

\paragraph{Fine-tuning strategies.} Let $\Theta$ denote all pretrained parameters and $\Phi$ the small task-specific parameter set. Depending on data budget and desired stability, we use one of the following:
\begin{itemize}
  \item \textbf{GNN-fixed (Output-only):} We freeze the modality encoders and the entire GNN backbone ($\Theta$) and train only the task-specific output heads ($\Phi$) on the new data. This preserves the shared representation capabilities while adapting the final mapping to the new label space.
  \item \textbf{Full-model fine-tuning:} Unfreeze $\Theta$ and optimize all parameters. This maximizes capacity but there is a risk of overfitting or drifting away from the learned shared representation when datasets are small.
  \item \textbf{Adaptor-based fine-tuning:} Keep $\Theta$ frozen and add small residual adaptor layers in the output head. Only the adaptors and a small subset of per-task scalars are trained, yielding fast training.
\end{itemize}

\subsection{Patent-derived fine-tuning example: UV-absorber formulation}
We evaluate an application-specific scenario by fine-tuning the base model on a compact, domain-focused dataset derived from a UV-absorber patent \citep{JP2004520284A}. The domain is the stabilization of organic materials. The concrete objective is reducing color fade in dyed wax under UV exposure. Each experiment is represented as our directed-tree process graph and provides multi-modal evidence comprising the stabilizer identity (novel benzotriazole derivatives from the patent or a conventional control) encoded as SMILES and embedded by the molecular encoder, the process conditions including additive loading (mass fraction) and the duration of exposure represented as numeric/value nodes with units, and the measured property, the color-difference magnitude $\Delta E$—a standard International Commission on Illumination (CIE) metric of total color change—used as the scalar target, where lower $\Delta E$ indicates better stabilization performance.

\begin{figure}[t]
\centering
\includegraphics[width=0.65\linewidth]{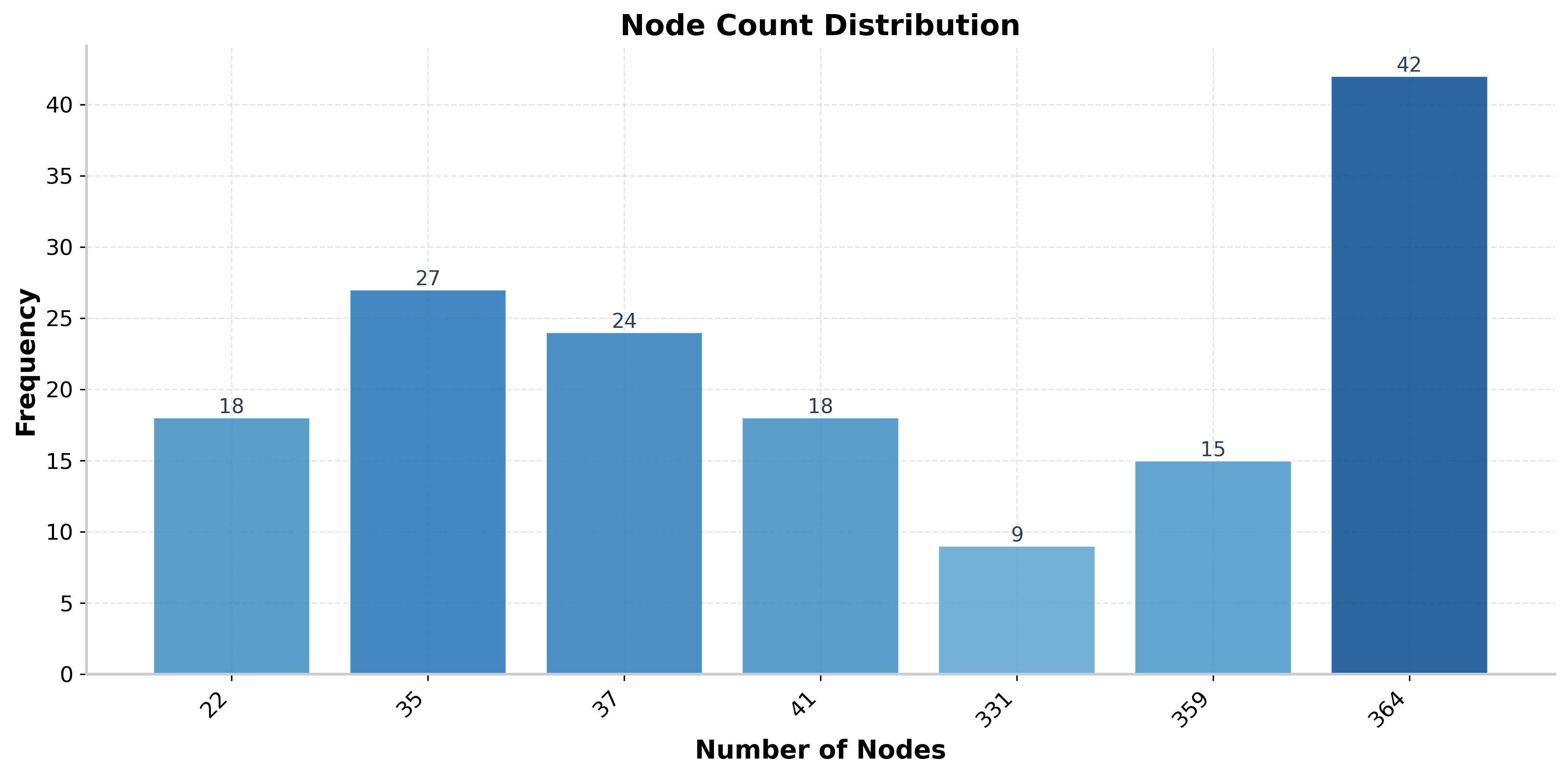}
\caption{Node count distribution in the directed-tree graphs of the UV-absorber fine-tuning dataset. The 153 process graphs span from compact formulations (22 nodes) to detailed multi-component processes (364 nodes). This structural diversity reflects the range of experimental descriptions captured in the patent and demonstrates the flexibility of our directed-tree representation.}
\label{fig:node_distribution}
\end{figure}

The dataset comprises 153 process graphs with diverse structural complexity, as shown in Figure~\ref{fig:node_distribution}. The node count distribution spans from compact formulations (22 nodes) to detailed multi-component processes (up to 364 nodes), reflecting the range of experimental descriptions captured in the patent. This structural diversity demonstrates that our representation can encode both simple screening experiments and richer process specifications within a single unified schema, and that the fine-tuned model must learn across this heterogeneity. Crucially, such variable-size graphs with differing numbers of nodes cannot be directly handled by classical machine learning methods that require fixed-length feature vectors; the graph neural network architecture naturally accommodates this structural variability through its message-passing and pooling operations, enabling end-to-end learning on heterogeneous process descriptions without manual feature engineering or padding artifacts. While it is common in molecular property prediction to flatten structures into fixed-length fingerprints (e.g., Morgan fingerprints), applying this approach to entire chemical processes incurs severe semantic loss. A process graph encodes causal dependencies---such as the specific ordering of steps and the local assignment of conditions to specific materials that are destroyed when flattened into global feature vectors (e.g., average temperature or bag-of-materials). Consequently, a comparison with traditional baselines (e.g., XGBoost) would primarily evaluate the quality of arbitrary manual feature engineering rather than the intrinsic predictability of the data. By contrast, our model learns directly from the ordered topological structure, preserving the distinction between identical operations performed at different stages of the workflow.

Fine-tuning proceeds by initializing from the pretrained backbone, instantiating a task head for the (patent, $\Delta E$) prediction, and optimizing only light-weight head/adaptor parameters (Appendix~\ref{app:adaptor_finetune_details}) with the heteroscedastic loss in Eq.~\eqref{eq:sigma_loss}. This setup preserves the universal process representation while adapting the final mapping to the domain-specific target. The case study illustrates how our representation naturally unifies molecular identity, formulation conditions, and performance labels, enabling efficient transfer from the base model to an industrially relevant prediction task. 

\begin{figure*}[t]
\centering
\begin{subfigure}[t]{0.48\textwidth} 
\centering
\includegraphics[width=\linewidth]{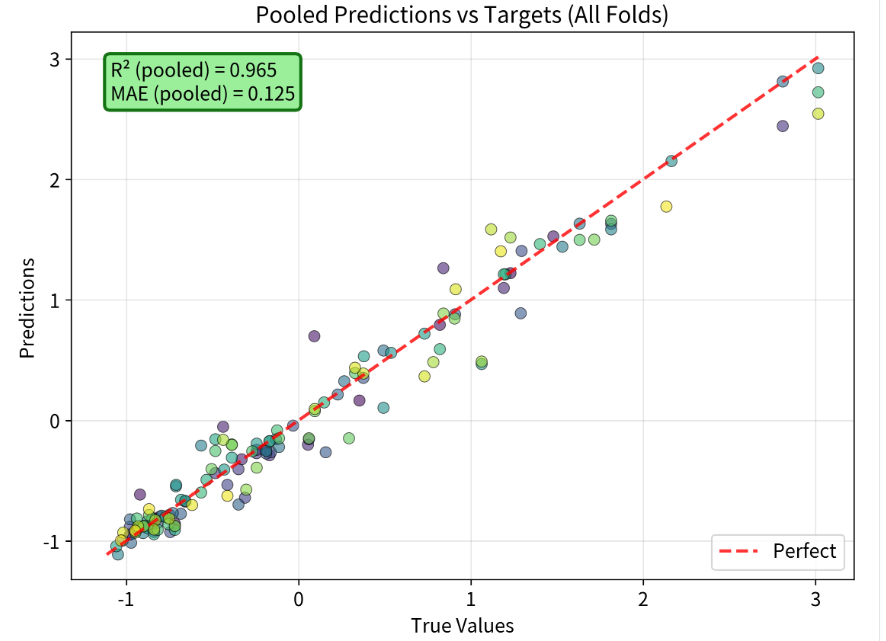}
\subcaption{GNN-fixed}
\end{subfigure}\hfill

\par\medskip 

\begin{subfigure}[t]{0.48\textwidth}
\centering
\includegraphics[width=\linewidth]{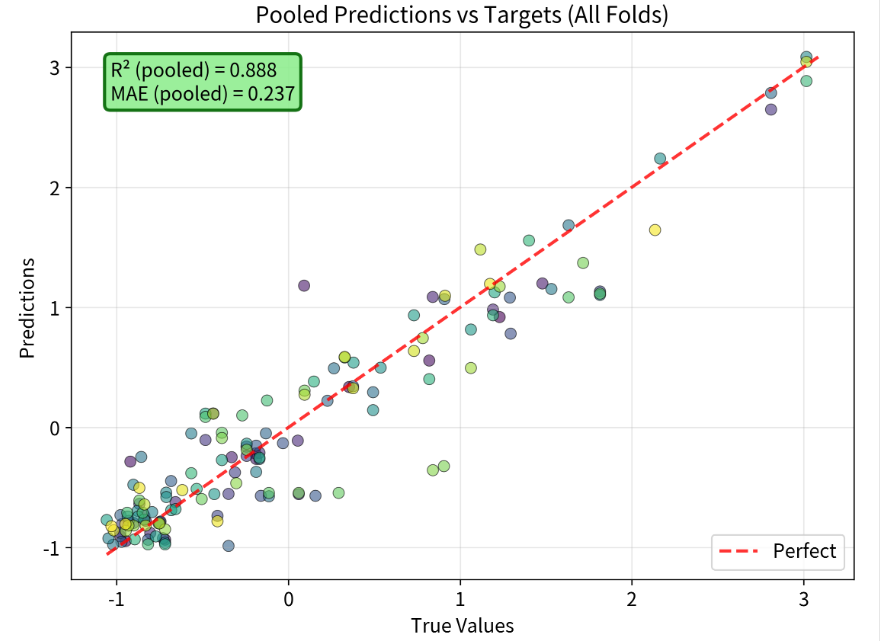}
\subcaption{Adaptor}
\end{subfigure}\hfill
\begin{subfigure}[t]{0.48\textwidth}
\centering
\includegraphics[width=\linewidth]{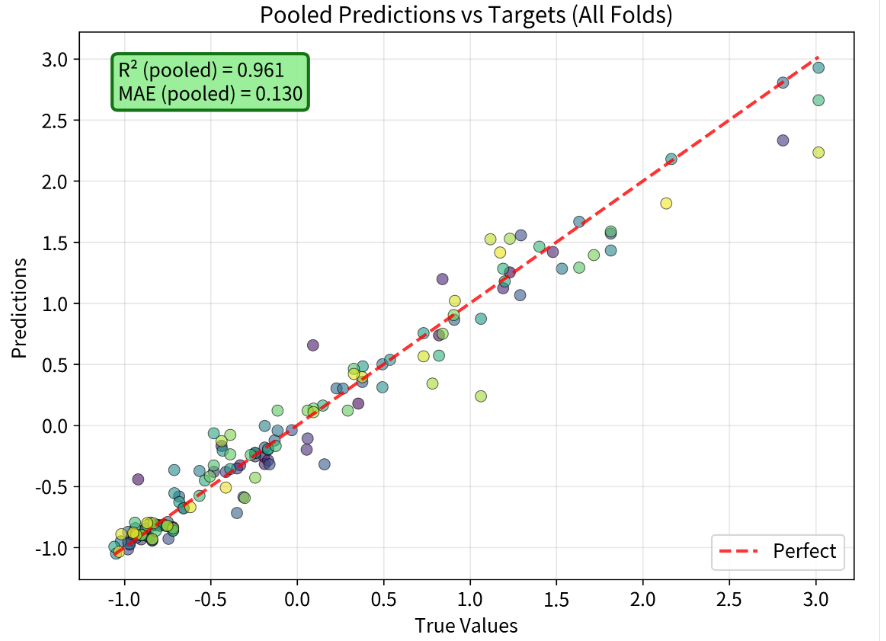}
\subcaption{Full-parameter}
\end{subfigure}
\caption{Predicted versus true values under three fine-tuning regimes on the UV-absorber task: (a) GNN-fixed, (b) adaptor, and (c) full-parameter. Points are pooled across cross-validation folds; the red dashed line indicates the perfect predictions, points of the same color belong to the same cross-validation fold.}
\label{fig:true_vs_predicted}
\end{figure*}

\begin{figure*}[t]
\centering
\begin{subfigure}[t]{0.48\textwidth}
\centering
\includegraphics[width=\linewidth]{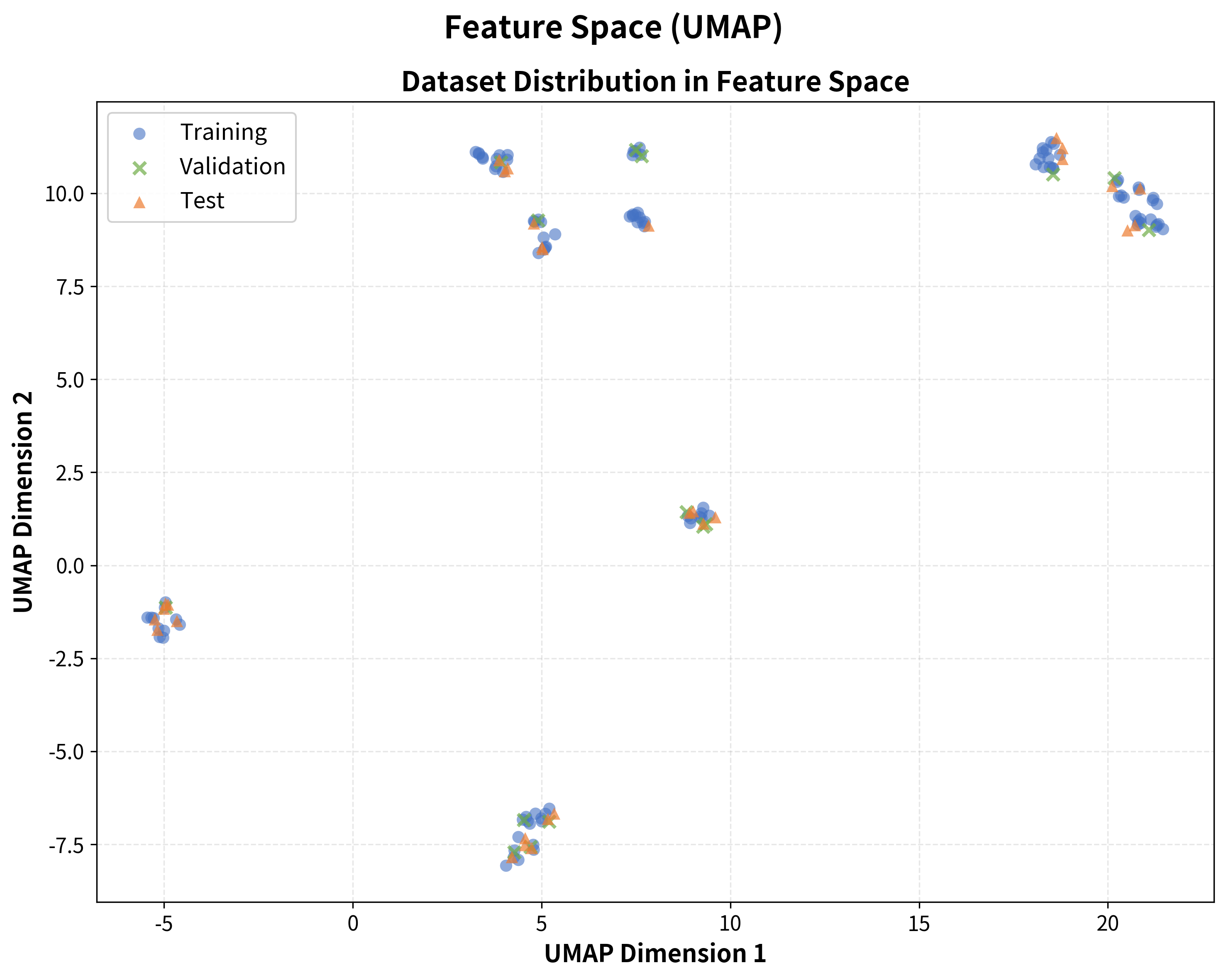}
\subcaption{GNN-fixed}
\end{subfigure}\hfill

\par\medskip

\begin{subfigure}[t]{0.48\textwidth}
\centering
\includegraphics[width=\linewidth]{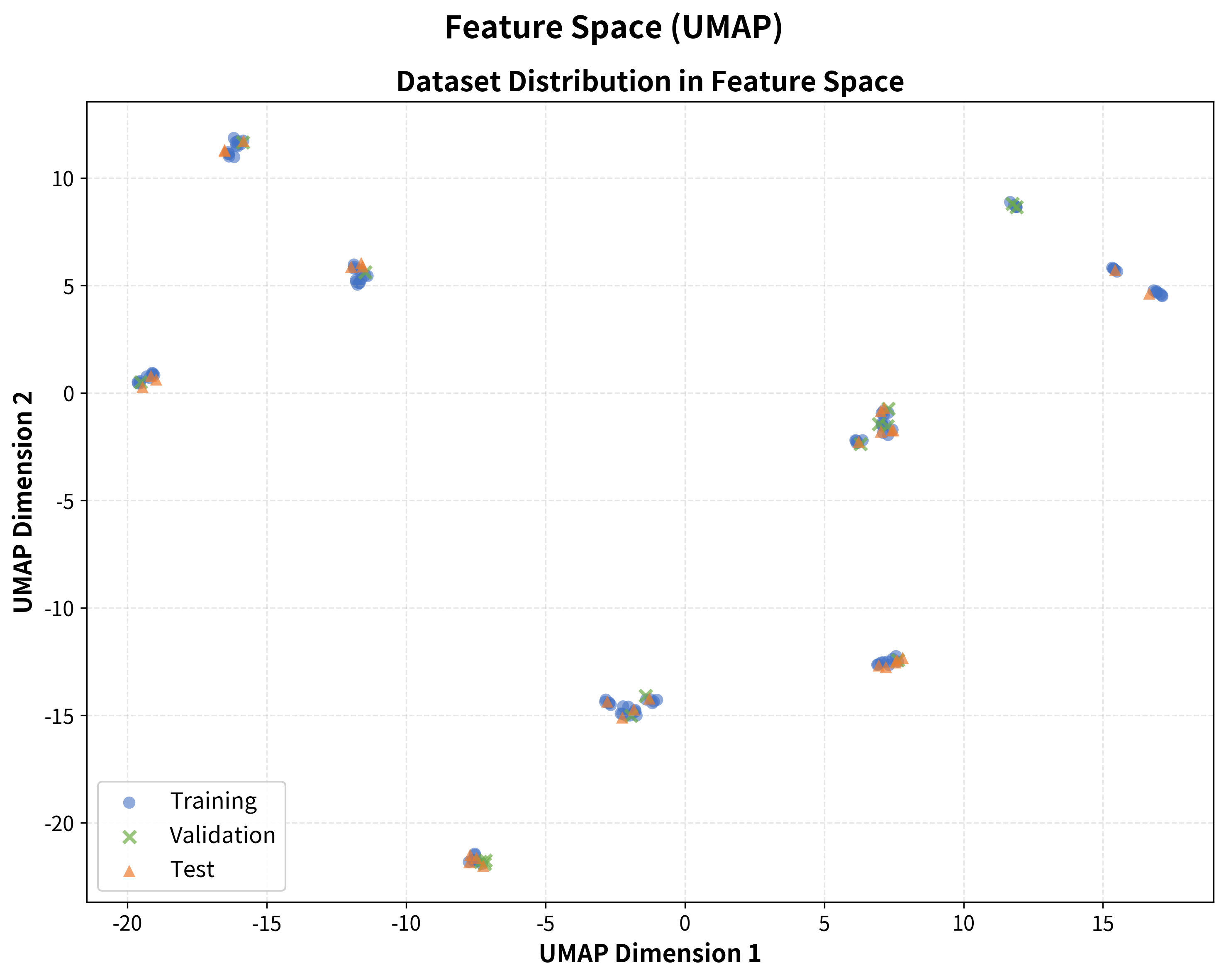}
\subcaption{Adaptor}
\end{subfigure}\hfill
\begin{subfigure}[t]{0.48\textwidth} 
\centering
\includegraphics[width=\linewidth]{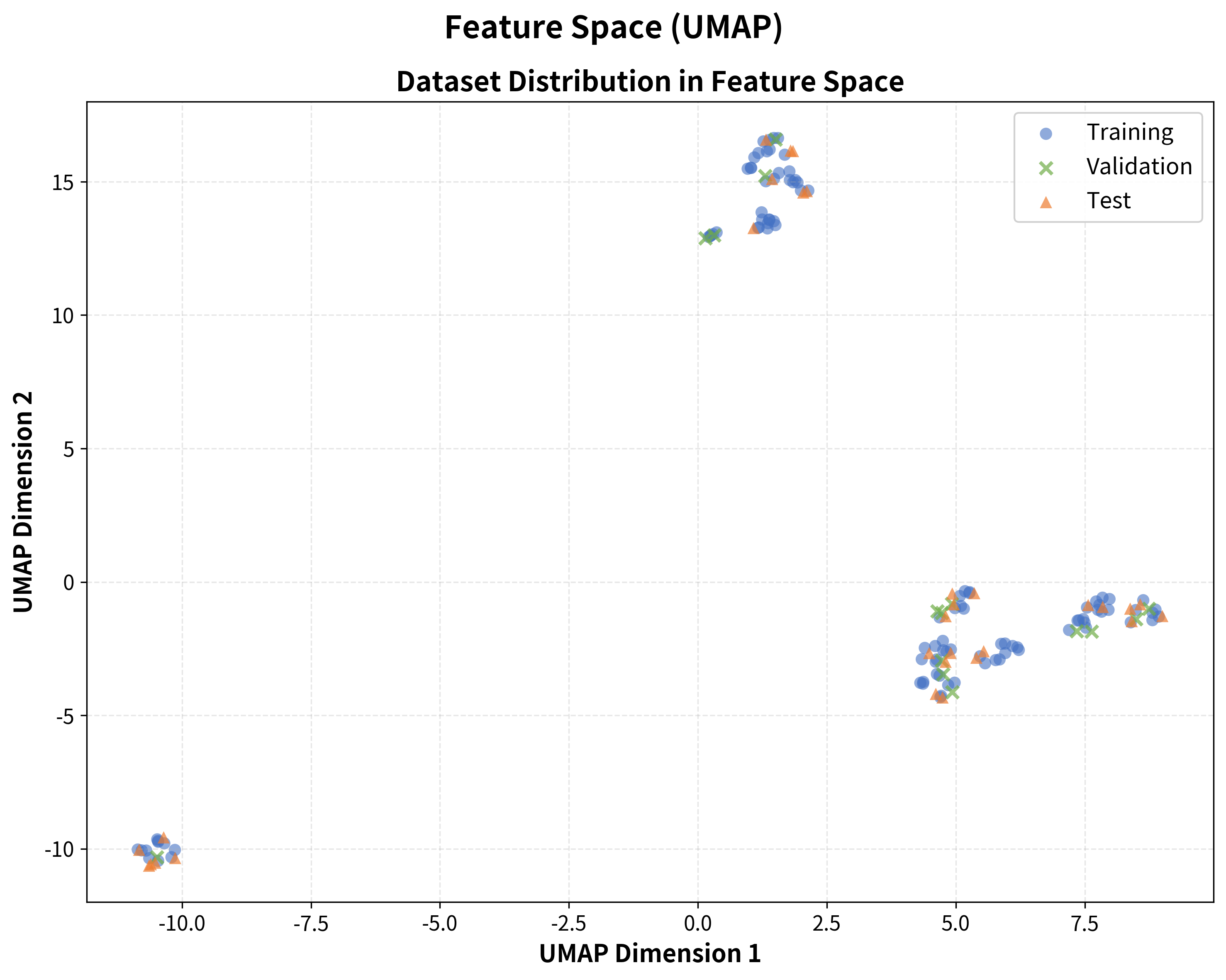}
\subcaption{Full-parameter}
\end{subfigure}
\caption{UMAP of property-conditioned latent embeddings under three fine-tuning regimes: (a) GNN-fixed, (b) adaptor, and (c) full-parameter. Points are colored by dataset split (training, validation, test).}
\label{fig:umap_features}
\end{figure*}

We evaluated the fine-tuned model using 5-fold cross-validation, with each fold using a 70\%/10\%/20\% split for training, validation, and test sets, respectively. Performance metrics reported in Table~\ref{tab:finetune_results} are computed by pooling predictions across all folds. The model achieves strong predictive performance with a pooled cross-validation $R^2$ score of 0.96, a mean absolute error (MAE) of 0.14, and a root mean squared error (RMSE) of 0.20 on normalized targets. These results demonstrate that the pretrained backbone successfully transfers to the domain-specific task with minimal additional training, confirming the effectiveness of our universal process representation for practical fine-tuning scenarios. In addition to pooled metrics, Figure~\ref{fig:true_vs_predicted} summarizes point-wise agreement between predictions and ground truth, while Figure~\ref{fig:umap_features} examines how the learned latent features distribute across train/val/test splits.

Figure~\ref{fig:true_vs_predicted} visualizes the predicted versus true property values across all five cross-validation folds. The scatter plot shows excellent agreement between predictions and ground truth, with data points tightly distributed along the diagonal. The pooled $R^2$ of 0.960 and MAE of 0.136 indicate that the fine-tuned model accurately captures the relationship between process parameters and the target color-difference metric across the entire dataset, including both training and validation samples. Reading the panels in order (a,b,c)—GNN-fixed, adaptor, and full-parameter—clarifies the capacity–stability trade-off: (a) freezing the backbone while training an expressive head yields the tightest clustering and the smallest error cloud; (b) the adaptor head underfits and shows the largest spread around the diagonal; (c) full-parameter fine-tuning remains competitive but exhibits a slightly wider dispersion than (a).

The JSON files with the human-curated experiments and their extracted, model-ready representations is provided in the Supplementary Information.

\begin{table}[!t]
\centering
{\setlength{\arrayrulewidth}{0.2pt}\arrayrulecolor{gray!30}%
\resizebox{\linewidth}{!}{%
\begin{tabular}{l c c c c c c} 
\toprule
\textbf{Regime} & \textbf{$R^2$ (pooled)} & \textbf{MAE (pooled)} & \textbf{RMSE (pooled)} & \textbf{\shortstack{$R^2$ mean\\$\pm$ std}} & \textbf{\shortstack{MAE mean\\$\pm$ std}} & \textbf{\shortstack{RMSE mean\\$\pm$ std}} \\
\midrule
Adaptor & 0.8880 & 0.2374 & 0.3301 & 0.8848 $\pm$ 0.0365 & 0.2374 $\pm$ 0.0417 & 0.3267 $\pm$ 0.0477 \\
\rowcolor{gray!10}\textbf{GNN-fixed} & \textbf{0.9648} & \textbf{0.1252} & \textbf{0.1851} & \textbf{0.9639 $\pm$ 0.0141} & \textbf{0.1252 $\pm$ 0.0273} & \textbf{0.1822 $\pm$ 0.0327} \\ 
Full-parameter & 0.9599 & 0.1362 & 0.1976 & 0.9596 $\pm$ 0.0177 & 0.1362 $\pm$ 0.0313 & 0.1928 $\pm$ 0.0436 \\ 
\bottomrule
\end{tabular}}%
}\arrayrulecolor{black}
\caption{Fine-tuning performance across regimes on the UV-absorber color-difference task (normalized targets). We report pooled cross-validation metrics and per-fold mean±std over 5 folds.}
\label{tab:finetune_results}
\end{table}
For the given example with only 153 samples, freezing the pretrained backbone while training an expressive head offers the best balance: it prevents overfitting to the small dataset (reducing variance) while the pretrained features provide sufficient inductive bias to fit the domain. However, we anticipate that for larger domain-specific datasets, full-parameter fine-tuning would likely become necessary to fully adapt the backbone representations. In our adaptor setup, the base is frozen and only two small residual adaptors plus per-task scale/bias are trained; the main projection weights remain effectively fixed. This limited capacity underfits the distribution shift, yielding lower $R^2$ and higher errors. By contrast, the GNN-fixed regime maintains stability of the backbone while allowing sufficient flexibility in the output projection to fit the domain, hence the best pooled metrics. 

\begin{figure}[htbp]
\centering
\includegraphics[width=0.78\linewidth,trim=5pt 2pt 5pt 8pt,clip]{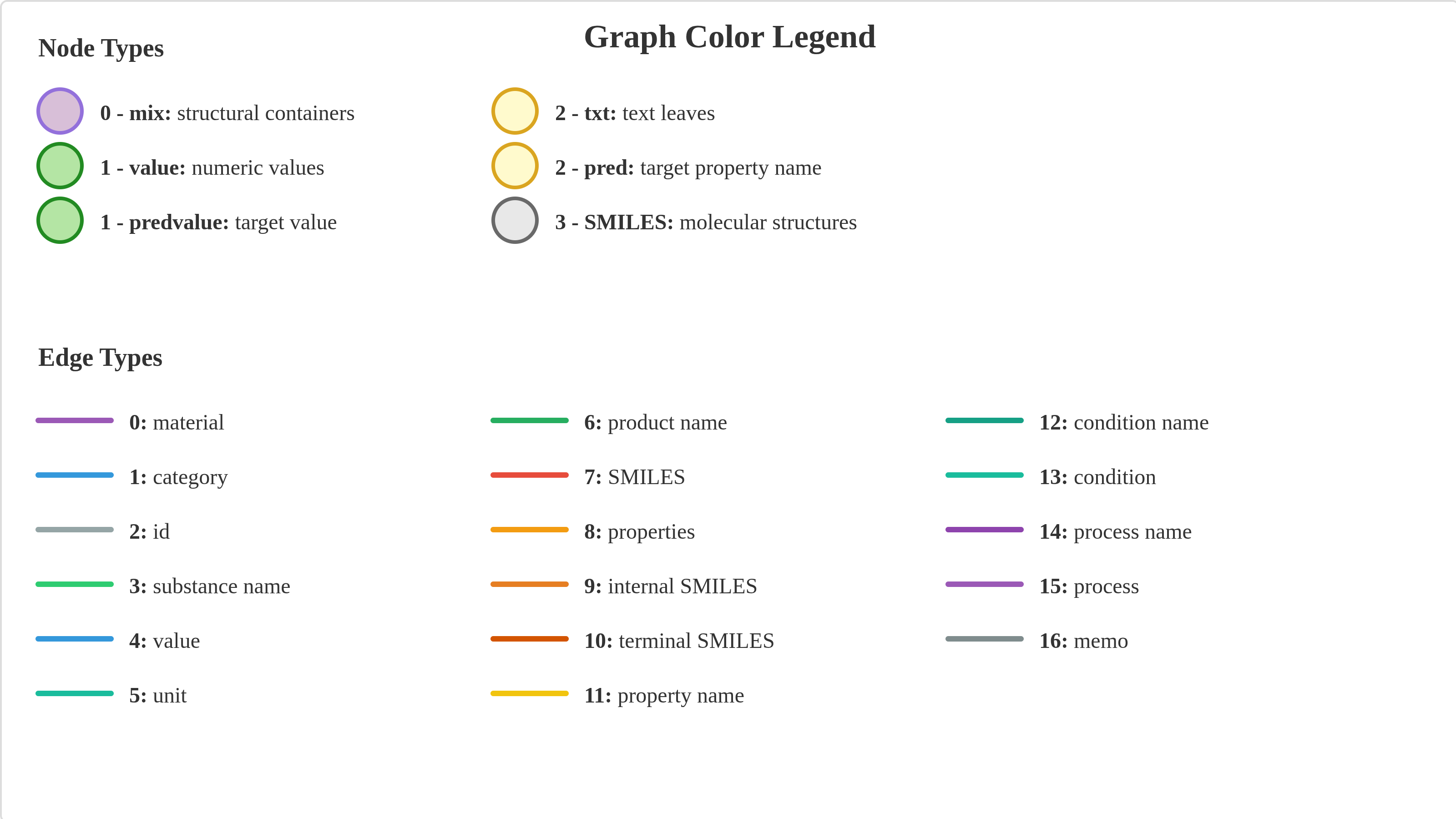}
\vspace{-1.0em}

\vspace{-0.9em}
\centering
\includegraphics[angle=-90,origin=c,width=0.98\linewidth,height=0.30\textheight,keepaspectratio]{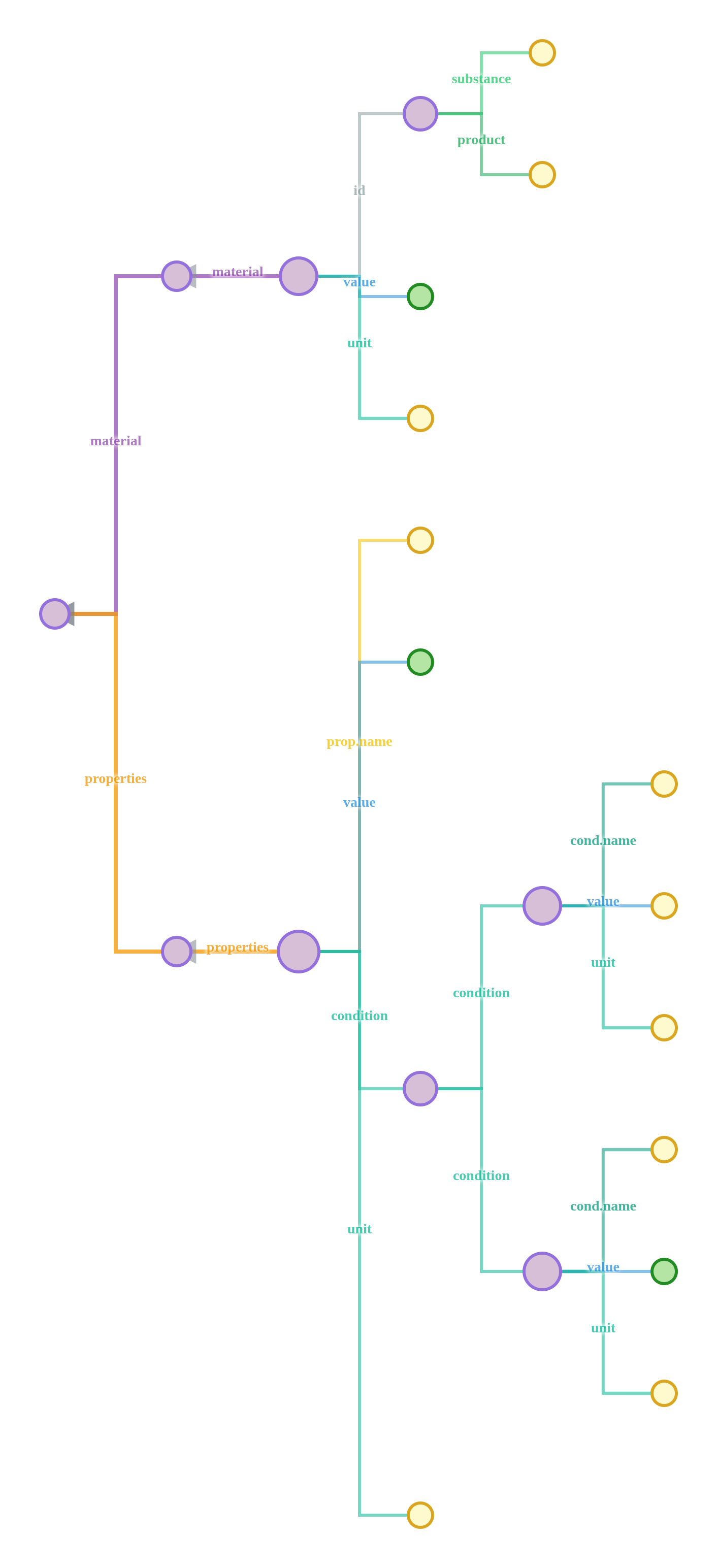}

\vspace{-5.6em}

\includegraphics[angle=-90,origin=c,width=0.98\linewidth,height=0.30\textheight,keepaspectratio]{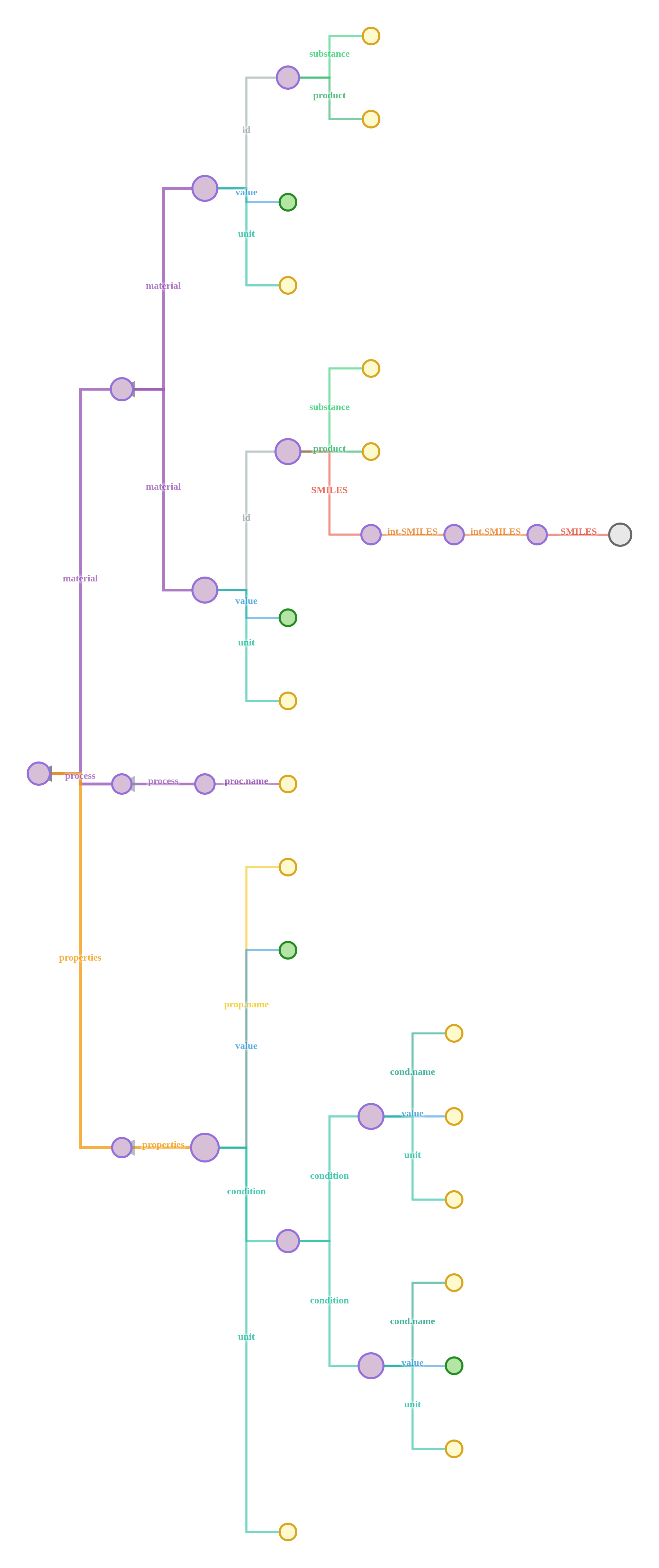}

\vspace{-8.1em}
\caption{Two representative directed-tree inputs from the UV-absorber study. The examples depict the chemical processes used in the experiments—materials, formulation conditions, and the resulting property.}
\label{fig:finetune_legend_examples}
\end{figure}

Figure~\ref{fig:finetune_legend_examples} visualizes this setup: the top panel is a color legend for node and edge types, and the two directed-tree inputs below are representative experiments from the UV-absorber study, summarizing materials, formulation conditions (e.g., loading and exposure duration), and the target color-change property. In total, the dataset comprises 7 types of graphs (in agreement with Figure~\ref{fig:node_distribution}). We provide only 2 graphs due to limited space; the other types of graphs are provided in the Supplementary Information.  

\FloatBarrier

\paragraph{Learned representations in latent space.} To understand how the model learns from chemical processes, we analyzed the learned latent representations using UMAP (Uniform Manifold Approximation and Projection)~\citep{McInnes2018UMAPUM}, a dimensionality-reduction technique that tries to preserve both global and local structures when projecting to two dimensions. We apply UMAP to the model's property-conditioned graph-level readout embeddings after message passing and attention pooling. UMAP constructs a $k$-nearest-neighbor graph in the original feature space (Euclidean metric) and embeds the concatenated train/validation/test features into a shared 2D space. We configured the projection with 15 nearest neighbors and a minimum distance of 0.1 to balance local neighborhood preservation with global structure, ensuring that the relative placement of experiments is directly comparable. Figure~\ref{fig:umap_features} visualizes these embeddings for the UV-absorber dataset; each point is a single experiment.

The UMAP projection reveals that the model learns a structured latent space organized into distinct clusters, where each cluster groups similar chemical processes sharing comparable materials, conditions, or outcomes. Crucially, training, validation, and test samples are well-mixed within these clusters rather than segregated, indicating that the model generalizes across data splits and captures the underlying physics rather than memorizing training examples. This structure emerges naturally from the multi-modal graph neural network, which maps diverse discrete inputs into a cohesive representation. While the global space is partitioned by formulation type, the local continuity within clusters suggests that interpolation between known conditions is meaningful, potentially allowing process optimization and active learning in the neighborhood of existing experiments.

Full-parameter fine-tuning yields the most flexible representation: clusters corresponding to related process graphs partially merge, reflecting greater manifold continuity and facilitating shared information transfer across nearby conditions. The GNN-fixed regime also supports significant information sharing, exhibiting distinct cluster mixing while preserving the underlying backbone geometry. In contrast, the adaptor regime produces the most rigid embedding—clusters remain compact and well separated—indicating limited information flow between different formulation types due to the restricted capacity of the adaptor head. Importantly, the three panels in Figure~\ref{fig:umap_features} show that training, validation, and test points occupy overlapping regions, indicating that our splits probe similar parts of the learned manifold and that the model generalizes beyond the training set.

\paragraph{Error behavior.} To contextualize the above metrics, we also examined per-regime residual distributions across folds; they are approximately centered and near-Gaussian with no heavy tails (Appendix Fig.~\ref{fig:residuals_distribution}), supporting the conclusions drawn from Table~\ref{tab:finetune_results} and Figure~\ref{fig:true_vs_predicted}. 

\section{Future Directions and Multi-Modal Extensions}
\label{sec:future_directions}
To further enhance calibration and sample efficiency for autonomous experimental planning, we plan to integrate the architecture with Gaussian process (GP)-based methods~\citep{Tsitsvero2025Learning}, specifically deep kernel learning on the shared latent and deep Gaussian processes atop the readout, as well as ensemble methods and Bayesian neural networks in the output heads. By leveraging the property-conditioned tokens for interpretable constraint selection, these probabilistic extensions will provide the robust uncertainty quantification required for practical closed-loop discovery. 

The flexibility of our directed-tree representation allows for natural extensions to include richer data modalities beyond the current text, numeric, and 2D molecular graph inputs. We envision extending the schema to incorporate 3D structural data, such as crystallographic information files (CIFs) from X-ray diffraction (XRD) or computed conformations, as well as spectral data like NMR or IR signals. These can be encoded via specialized geometric GNNs or spectral encoders and attached as attributes to material nodes, providing the model with explicit structural and electronic context that 2D molecular graphs cannot fully capture.

Visual data is also central to materials characterization. The graph structure can readily accommodate image nodes representing particle morphology (e.g., from scanning or transmission electron microscopy, SEM/TEM) or pore distributions, which are critical in domains like additive manufacturing and catalysis. Similarly, for device-level modeling, blueprints or schematic images (e.g., battery cell configurations) can be embedded and linked to process steps, allowing the model to correlate fabrication geometry with performance.

Finally, integrating this predictive framework with generative AI opens a path to holistic materials design. Generative models can propose novel 3D molecular or crystal structures targeted for specific properties. These generated candidates can then be inserted into our process-graph representation as input materials to evaluate their feasibility and performance within realistic manufacturing workflows. This synergy, generating candidates and virtually screening them in the context of their synthesis and processing, moves beyond simple structure-property prediction toward comprehensive experimental design.

\section{Conclusion}
\label{sec:conclusion}
We have presented a framework for machine-learning-driven prediction of material properties that addresses the challenge of learning from heterogeneous experimental data. The core contribution is a universal directed-tree representation that unifies chemical structures (SMILES), text, and numerical values into a single machine-readable format suitable for graph neural networks. To learn from this representation, we developed a multi-modal GNN with property-conditioned attention that was trained on approximately 700,000 process graphs derived from nearly 9,000 diverse documents spanning polymers, electronics, energy materials, and industrial formulations.

We demonstrated the framework's practical utility through fine-tuning experiments on a UV-absorber formulation task. Using only 153 domain-specific samples, the pretrained backbone achieved an $R^2$ of 0.96, validating that the learned representations transfer effectively to specialized prediction tasks. The GNN-fixed fine-tuning strategy proved most effective for small datasets by preventing overfitting while leveraging the pretrained features.

A current limitation is the reliance on human expert curation of process graphs. Going forward, we will automate this pipeline using LLM-assisted extraction with structured parsing to improve scalability. Future work will also include rigorous benchmarking against traditional fingerprint-based baselines to quantify the specific predictive gains of graph-based process representations.

\section*{Author Contributions}
Atsuyuki Nakao developed the data format for representing chemical processes and designed the base model architecture. Mikhail Tsitsvero and Atsuyuki Nakao performed base model training and designed the fine-tuning experiments. Mikhail Tsitsvero and Atsuyuki Nakao prepared the manuscript. Atsuyuki Nakao, Mikhail Tsitsvero, and Hisaki Ikebata prepared the data for model training. Atsuyuki Nakao, Mikhail Tsitsvero, and Hisaki Ikebata reviewed and discussed the manuscript.

\section*{Acknowledgments}
This paper is based on results obtained from a project, JPNP23019, subsidized by the New Energy and Industrial Technology Development Organization (NEDO), Japan. 

\newpage

\bibliographystyle{iclr2026_conference}
\bibliography{bibliography}
 

\newpage

\appendix 
\section{Appendix} 

\subsection{Loss function}
\label{app:heteroscedastic_loss}
Starting from the standard mean squared error (MSE) objective, $\frac{1}{N}\sum_{i=1}^{N}(y_i{-}p_i)^2$, assume Gaussian observation noise with task-specific variance $\sigma_{t(i)}^2$ for sample $i$ belonging to task $t(i)$. Maximizing the Gaussian likelihood is equivalent to minimizing the negative log-likelihood
\[
\frac{1}{N}\sum_{i=1}^{N}\left(\frac{(y_i - p_i)^2}{2\,\sigma_{t(i)}^2} + \tfrac{1}{2}\log\big(2\pi\,\sigma_{t(i)}^2\big)\right),
\]
which yields an MSE term reweighted by the inverse variance plus a log-variance penalty. Promoting $\sigma_{t}$ to learned per-task parameters gives a simple heteroscedastic variant of MSE. We also include a small $\ell_2$ penalty on the head's task-modulation vector $\mathbf{w}_{t(i)}$ to regularize the task-specific scaling. With a numerical stabilizer $\varepsilon{=}10^{-2}$, the loss used in our experiments is
\begin{equation}
\label{eq:sigma_loss}
\mathcal{L}_{\sigma} \,=\, \underbrace{\,\frac{1}{N}\sum_{i=1}^{N}\log\big(2\pi\,(\sigma_{t(i)}{+}\varepsilon)^2\big)\,}_{\text{log-variance term}} \;+
\; \underbrace{\,\frac{1}{N}\sum_{i=1}^{N}\frac{(y_i - p_i)^2}{(\sigma_{t(i)}{+}\varepsilon)^2}\,}_{\text{variance-normalized SE}} \;+
\; \underbrace{\,\lambda\,\frac{1}{N}\sum_{i=1}^{N}\lVert \mathbf{w}_{t(i)} \rVert_2^2\,}_{\text{output-weight regularizer}}\; ,
\end{equation}
where $\sigma_{t(i)}$ is the learned per-task standard deviation for task $t(i)$, $\mathbf{w}_{t(i)}$ is the task-specific modulation vector from Eq.~\eqref{eq:pred_modulation}, and $\lambda$ is a regularization coefficient. In our implementation, the first two terms are computed by a dedicated loss module, while the regularizer is added separately in the training loop. All losses are computed on normalized targets. Both $\sigma_{t(i)}$ and $\mathbf{w}_{t(i)}$ are stored in task-indexed embedding tables, enabling efficient parameter sharing across the approximately 50,000 (document, property) tasks in our dataset. In practice we use decoupled weight decay (AdamW-style) on trainable parameters in addition to the $\ell_2$ term above; the per-task variance parameters are learned jointly with the prediction heads.

\subsection{Base-model training details}
\label{app:base_training_details}

\paragraph{Computational costs and training time.} The base model was trained on a cluster equipped with $8{\times}$NVIDIA A100 (80\,GB VRAM) GPUs using distributed data-parallel (DDP) training. End-to-end pretraining on the full dataset of approximately 700,000 process graphs completed in approximately 6 hours (wall-clock time). Training was performed until the training $R^2$ score reached approximately $0.75$ on all available data, at which point the model had converged and further training yielded diminishing returns on validation set (evaluated on the previous train/val/test run). This relatively modest training time demonstrates the practical feasibility of the approach for research and industrial settings.

\paragraph{Optimization and hyperparameters.} For both pretraining and fine-tuning we used the schedule-free optimizer~\citep{defazio2024schedule} with an initial learning rate of $10^{-3}$, together with light decoupled weight decay on all trainable parameters. Per-task variance parameters were learned jointly with the heads as described in Appendix~\ref{app:heteroscedastic_loss}. Unless otherwise noted, all metrics are reported on normalized targets.

\subsection{Adaptor-based fine-tuning details}
\label{app:adaptor_finetune_details}
We use parameter-efficient residual adaptors in the output head to adapt the pretrained backbone without modifying shared parameters. Denote the graph-level latent $\mathbf{z}_{\text{readout},i}\in\mathbb{R}^{d}$ produced by the frozen backbone for sample $i$, and a task index $t(i)$. The standard head produces
\[\tilde{\mathbf{z}}_i = \tanh\!\big(\mathrm{LN}(\mathbf{W}_z(\mathbf{y}_{\text{prop}})\,\mathbf{z}_{\text{readout},i}+\mathbf{b}_z(\mathbf{y}_{\text{prop}}))\big),\]
\[p_i= a_{t(i)}\, \Big(\big(\mathbf{W}_p(\mathbf{y}_{\text{prop}}) \odot (1 + \tfrac{\mathbf{w}_{t(i)}}{100})\big)^{\top}\tilde{\mathbf{z}}_i\Big) + b_{t(i)},\]
where $\mathbf{W}_z(\cdot)$, $\mathbf{b}_z(\cdot)$, and $\mathbf{W}_p(\cdot)$ are shared property-conditioned projections derived from the target property embedding $\mathbf{y}_{\text{prop}}$, and $a_{t}, b_{t}, \mathbf{w}_t$ are per-task parameters (cf. Eq.~\eqref{eq:pred_modulation}).

\paragraph{Training configuration.} For the UV-absorber fine-tuning experiments reported in Section~\ref{sec:finetuning}, we used a schedule-free optimizer~\citep{defazio2024schedule} with an initial learning rate of $10^{-3}$ and trained for 20 epochs. The schedule-free formulation eliminates the need for manual learning-rate schedules while maintaining fast convergence and stable adaptation of the adaptor parameters.

Adaptor fine-tuning augments two residual terms while freezing the shared weights: (i) an input-side adaptor \(\mathbf{A}_1\in\mathbb{R}^{d\times d'}\) that maps the readout latent to an additive residual in the model dimension, and (ii) an output-side adaptor \(\mathbf{A}_2\in\mathbb{R}^{d\times 1}\) providing a small residual to the scalar prediction. In our implementation, the input-side adaptor is applied directly to the readout vector $\mathbf{z}_{\text{readout},i}$ (i.e., $d' = d_{\text{readout}}$, the dimension of the graph-level latent). Concretely,
\[\hat{\mathbf{z}}_i = \tanh\!\Big(\mathrm{LN}\big(\mathbf{W}_z(\mathbf{y}_{\text{prop}})\,\mathbf{z}_{\text{readout},i} + \mathbf{b}_z(\mathbf{y}_{\text{prop}}) + \mathbf{A}_1\,\mathbf{z}_{\text{readout},i}\big)\Big),\]
and the final prediction becomes
\[p_i\;=\; a_{t(i)}\,\Big(\big(\mathbf{W}_p(\mathbf{y}_{\text{prop}}) \odot (1 + \tfrac{\mathbf{w}_{t(i)}}{100})\big)^{\top}\hat{\mathbf{z}}_i\Big) + b_{t(i)}\; +\; \mathbf{A}_2^{\top}\,\hat{\mathbf{z}}_i.\]
In practice we zero-initialize $\mathbf{A}_1,\mathbf{A}_2$ so that the initial adaptor is an identity-respecting perturbation (no change at step 0), and optimize only $\{\mathbf{A}_1,\mathbf{A}_2, a_t, b_t\}$ while all backbone and shared head weights remain frozen. Training uses the same heteroscedastic objective as in the main text (Eq.~\eqref{eq:sigma_loss}), optionally with weight decay on adaptor parameters to regularize their magnitude. This yields stable adaptation with a parameter count that is a small fraction of the full model.

Variants mirror common PEFT designs: bottleneck adaptors (choose $d'\ll d$), per-task or shared adaptors, and selective unfreezing of the last $K$ graph layers in combination with adaptors for higher capacity when more data is available.

\subsection{Residual analysis of fine-tuning performance}
\label{app:residuals_analysis}
\begin{figure*}[t]
\centering
\begin{subfigure}[t]{0.48\textwidth}
\centering
\includegraphics[width=\linewidth]{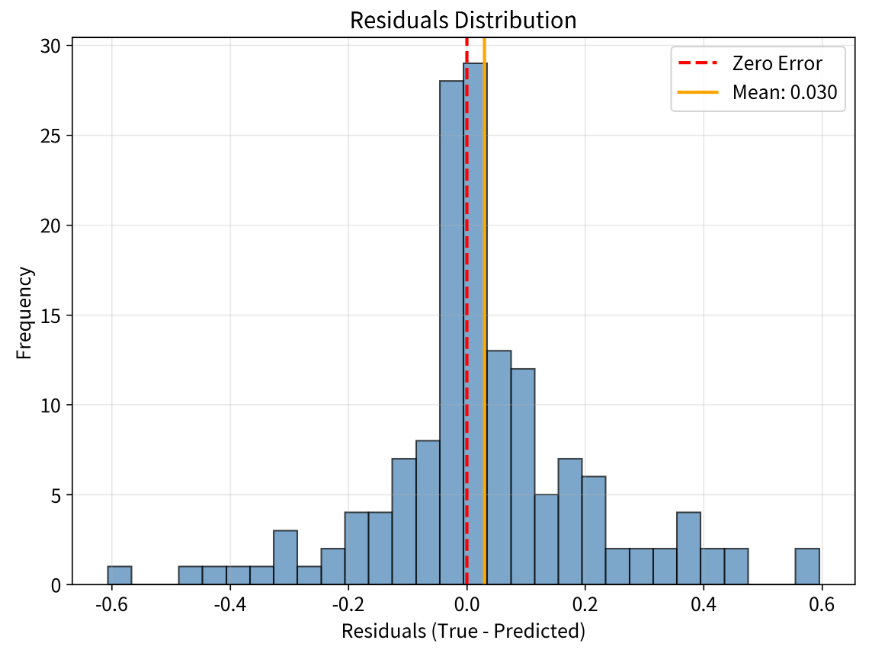}
\subcaption{GNN-fixed}
\end{subfigure}\hfill

\par\medskip

\begin{subfigure}[t]{0.48\textwidth}
\centering
\includegraphics[width=\linewidth]{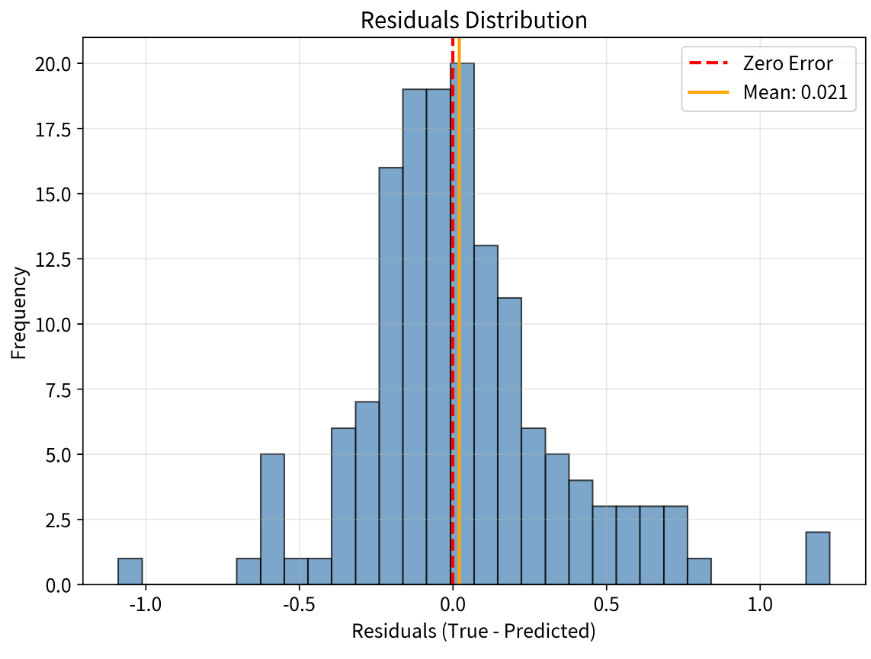}
\subcaption{Adaptor}
\end{subfigure}\hfill
\begin{subfigure}[t]{0.48\textwidth}
\centering
\includegraphics[width=\linewidth]{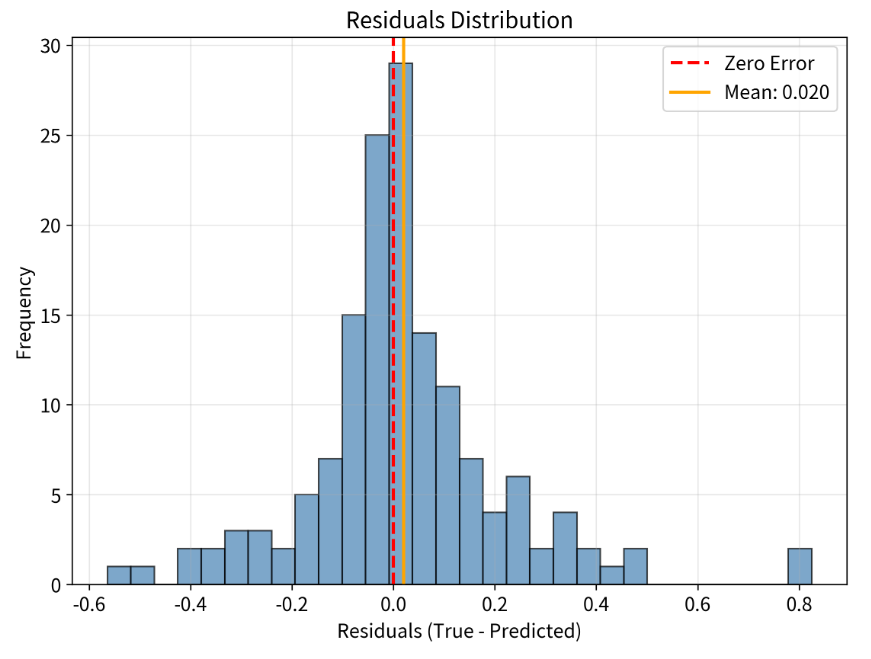}
\subcaption{Full-parameter}
\end{subfigure}
\caption{Residuals (true minus predicted) under three fine-tuning regimes on the UV-absorber task: (a) GNN-fixed, (b) adaptor, and (c) full-parameter. Histograms are pooled across cross-validation folds; the red dashed line marks zero error and the orange line the empirical mean.}
\label{fig:residuals_distribution}
\end{figure*}

To assess the quality and systematic behavior of prediction errors under each fine-tuning regime, we analyzed residuals (true minus predicted) pooled across all cross-validation folds for the UV-absorber task. As shown in Figure~\ref{fig:residuals_distribution}, all three regimes are approximately centered near zero with small positive means (${\approx}0.02$--$0.03$). The GNN-fixed model exhibits the narrowest spread, the adaptor head shows slightly wider tails, and the full-parameter model is comparable with a mild right tail. None display heavy tails or multimodality, consistent with the strong metrics in Table~\ref{tab:finetune_results}.

\subsection{Graph Neural Networks with Attention} 
\label{app:gnn_attention}
Graph neural networks (GNNs) operate by iteratively exchanging information among neighboring nodes via message passing. A generic layer can be written as
\[\mathbf{m}_i = \sum_{j\in\mathcal{N}(i)} \psi\big(\mathbf{h}_i,\,\mathbf{h}_j,\,\mathbf{e}_{ij}\big),\qquad
\mathbf{h}'_i = \phi\big(\mathbf{h}_i,\,\mathbf{m}_i\big),\]
where \(\mathbf{h}_i\) denotes node features, \(\mathbf{e}_{ij}\) optional edge features, \(\psi\) is a learnable message function, and \(\phi\) is an update function. Repeating this over layers lets information flow along the graph.

In this work we use a Transformer-style graph convolution layer implemented in PyTorch Geometric \citep{ijcai2021p214,velickovic2018graph,Fey/Lenssen/2019}. It performs multi-head dot-product attention over neighbors, optionally incorporating edge attributes. For one head, the layer can be expressed as
\[\mathbf{h}'_i = \mathbf{W}_1\,\mathbf{h}_i 
\; + \; \sum_{j\in\mathcal{N}(i)} \alpha_{ij}\,\Big(\mathbf{W}_2\,\mathbf{h}_j 
\; + \; \mathbf{W}_e\,\mathbf{e}_{ij}\Big),\]
with attention coefficients
\[\alpha_{ij} \;=\; \mathrm{softmax}_{j}\!\left(\frac{\big(\mathbf{W}_q\,\mathbf{h}_i\big)^{\top}\,\big(\mathbf{W}_k\,\mathbf{h}_j + \mathbf{W}_e\,\mathbf{e}_{ij}\big)}{\sqrt{d}}\right),\]
where $\mathbf{W}_q$ and $\mathbf{W}_k$ are query and key projections, respectively, and $\mathbf{W}_e$ projects edge attributes into the attention computation. When using multiple heads, outputs are concatenated or averaged. We used categorical edge-label embeddings as edge attributes so that attention keys/values and the final aggregation are edge-aware, following the reference implementation in PyTorch Geometric\,\citep{Fey/Lenssen/2019}.

\subsection{Property-conditioned readout} 
\label{app:prop_readout}
Let $\mathbf{X}=\big[\mathbf{x}_1,\ldots,\mathbf{x}_n\big]^{\top}\in\mathbb{R}^{n\times d_n}$ be the matrix of structural-node features after graph message passing, and let $\mathbf{y}_{\mathrm{prop}}\in\mathbb{R}^{d_t}$ denote the target property embedding. We use multi-head, property-conditioned attention both to build a small set of latent tokens and for the final readout. With $H$ heads, $d_q$ query/key and $d_v$ value dimensions per head, linear projections are
\[\mathbf{q}=\mathbf{W}_q\,\mathbf{y}_{\mathrm{prop}},\qquad \mathbf{K}=\mathbf{X}\,\mathbf{W}_k,\qquad \mathbf{V}=\mathbf{X}\,\mathbf{W}_v,\]
where $\mathbf{W}_q\in\mathbb{R}^{d_t\times Hd_q}$, $\mathbf{W}_k\in\mathbb{R}^{d_n\times Hd_q}$, and $\mathbf{W}_v\in\mathbb{R}^{d_n\times Hd_v}$ are learned (no bias). Writing per-head slices as $\mathbf{q}_h\in\mathbb{R}^{d_q}$, $\mathbf{k}_{jh}\in\mathbb{R}^{d_q}$, and $\mathbf{v}_{jh}\in\mathbb{R}^{d_v}$, attention weights and the head outputs are
\[\alpha_{jh} = \mathrm{softmax}_{j}\!\left( \frac{\mathbf{q}_h^{\top}\mathbf{k}_{jh}}{\sqrt{d_q}} \right),\qquad \mathbf{u}_h = \sum_{j=1}^n \alpha_{jh}\,\mathbf{v}_{jh}.\]
The aggregated vector is obtained by concatenating the per-head outputs
\[\mathbf{z}=\big[\mathbf{u}_1\,\|\,\cdots\,\|\,\mathbf{u}_H\big]\in\mathbb{R}^{Hd_v}.\]
Stacked selection layers use the same mechanism to map $n\!\to\!n'$ nodes (tokens), with residual connections between blocks. The final readout applies property-conditioned attention over the token set and returns a graph-level vector $\mathbf{z}_i \in \mathbb{R}^{d_z}$, which is used by the task-modulated prediction in \Eqref{eq:pred_modulation}.

\subsection{Dataset Curation and Scope}
\label{app:dataset_curation}
The training dataset was sourced from a comprehensive collection of nearly 9,000 patent documents and scientific articles. The chemical process information within these documents—including experimental procedures, material compositions, processing conditions, and measured properties—was systematically extracted and curated by human experts to ensure high fidelity and structural consistency. Each curated entry was then converted into our universal directed-tree representation.

The documents were sourced globally from multiple patent offices and scientific publishers. This collection includes a significant number of documents from the Japan Patent Office (JPO), as well as substantial contributions from the United States Patent and Trademark Office (USPTO), the World Intellectual Property Organization (WIPO), and the European Patent Office (EPO), among others. This geographical diversity ensures the model is trained on a broad range of experimental practices and reporting styles.

The dataset is intentionally diverse, covering a wide spectrum of industrial and research domains to ensure the model learns a truly generalizable representation of chemical processes. The domains include, but are not limited to:
\begin{itemize}
    \item \textbf{Polymers and Resins:} Engineering plastics, biodegradable and photosensitive resins, adhesives, elastomers, and packaging films.
    \item \textbf{Electronics and Semiconductors:} Materials for printed circuit boards, encapsulants, CMP polishing slurries, color filter resists, and conductive particles.
    \item \textbf{Energy and Environmental Materials:} Components for lithium-ion batteries, CO2 absorbents, and catalysts.
    \item \textbf{Advanced and Composite Materials:} Ceramics, carbon materials, glass, synthetic fibers, magnetic materials, and prepregs.
    \item \textbf{Industrial Formulations:} Automotive and solar-reflective paints, tire rubber, lubricating oils, inks, and cosmetics.
\end{itemize}
This breadth ensures that the model is exposed to a vast range of materials, conditions, and property relationships, which is foundational to the cross-domain transferability demonstrated in our fine-tuning experiments.

\end{document}

%% file: math_commands.tex
\usepackage{amsmath,amsfonts,bm}

\def\eqref#1{equation~\ref{#1}}
\def\Eqref#1{Equation~\ref{#1}}

\def\1{\bm{1}}

\DeclareMathAlphabet{\mathsfit}{\encodingdefault}{\sfdefault}{m}{sl}
\SetMathAlphabet{\mathsfit}{bold}{\encodingdefault}{\sfdefault}{bx}{n}

